\documentclass[11pt,a4paper]{article}
\pdfoutput=1
\usepackage{jcappub}
\usepackage{graphicx}
\usepackage{booktabs}

\newcommand{\bfeta}{\boldsymbol\eta}
\newcommand{\beqra}{\begin{eqnarray}}
\newcommand{\eeqra}{\end{eqnarray}}
\newcommand{\beq}{\begin{equation}}
\newcommand{\eeq}{\end{equation}}

\title{Global fits of the dark matter-nucleon effective interactions}
\author[a]{Riccardo Catena}
\author[b]{and Paolo Gondolo}
\affiliation[a]{Institut f\"ur Theoretische Physik, Friedrich-Hund-Platz 1, 37077 G\"ottingen, Germany}
\affiliation[b]{Department of Physics and Astronomy, University of Utah, 115 South 1400 East \#201, Salt Lake City, UT 84112, USA}
\emailAdd{riccardo.catena@theorie.physik.uni-goettingen.de}
\emailAdd{paolo.gondolo@utah.edu}

\abstract{ The effective theory of isoscalar dark matter--nucleon interactions mediated by heavy spin-one or spin-zero particles depends on 10 coupling constants besides the dark matter particle mass. Here we compare this 11-dimensional effective theory to current observations in a comprehensive statistical analysis of several direct detection experiments, including the recent LUX, SuperCDMS and CDMSlite results. From a multidimensional scan with about 3 million likelihood evaluations, we extract the marginalized posterior probability density functions (a Bayesian approach) and the profile likelihoods (a frequentist approach), as well as the associated credible regions and confidence levels, for each coupling constant vs dark matter mass and for each pair of coupling constants. We compare the Bayesian and frequentist approach in the light of the currently limited amount of data. We find that current direct detection data contain sufficient information to simultaneously constrain not only the familiar spin-independent and spin-dependent interactions, but also the remaining velocity and momentum dependent couplings predicted by the dark matter-nucleon effective theory. For current experiments associated with a null result, we find strong correlations between some pairs of coupling constants. For experiments that claim a signal (i.e., CoGeNT and DAMA), we find that pairs of coupling constants produce degenerate results.}


\begin{document}
\maketitle

\section{Introduction}
Only about one sixth of the total matter in the observable Universe is made of known particles~\cite{Ade:2013zuv}. The remaining part is an invisible and unidentified cosmic component called the dark matter~\cite{Zwicky,Kolb:1990vq,Jungman:1995df,Bertone:2004pz}. The particles forming the dark matter component have up to now escaped detection.  Astronomical observations and numerical simulations~\cite{Kuhlen:2012ft} show that dark matter clusters in large astrophysical structures (dark matter halos). The clustering of dark matter particles has inspired a number of complementary methods to detect them (see, e.g.,~\cite{Strigari:2013iaa} for a recent review).  In particular, dark matter particles in our own Milky Way galaxy can be searched for using the direct detection technique~\cite{Goodman:1984dc}, which has recently played an important role in this context. Several direct detection experiments have indeed reached the sensitivity to probe the dark matter paradigm (see, e.g.,~\cite{Baudis:2012ig} for a review). The goal of the direct detection technique is to measure the energy deposited in an underground detector by Milky Way dark matter particles scattering on a target material~\cite{Cerdeno:2010jj}. This detection strategy is therefore an ideal tool to probe the foundations of the dark matter-nucleon interaction.

There is an extensive literature devoted to the study of dark matter scattering on nuclei in an underground detector (for an overview of this subject, see for instance Refs.~\cite{DelNobile:2013gba,Frandsen:2013cna,Schwetz:2011xm,Farina:2011pw,Kopp:2009qt,Savage:2010tg} and references therein). The vast majority of these analyses rely on the assumption that dark matter interacts with the detector nuclei either through the nuclear charge density operator or through the nuclear spin-current density operator.  The former is commonly called ``spin-independent interaction,''  the latter  ``spin-dependent interaction,'' although strictly speaking any other operator is either spin-dependent or spin-independent. This approach to the dark matter direct detection is motivated by its simplicity and by the fact that these two interaction operators are naturally generated in the most popular dark matter models. Important examples are those based on the Minimal Supersymmetric Standard Model or on many of its extensions~\cite{Baltz:2004aw,Ellis:2005mb,Catena:2013pka}. On the other end, there is no empirical evidence supporting this assumption, and Nature might actually be more complex, allowing a broader spectrum of possible dark matter-nucleon interactions. 

In the past few years alternative types of dark matter-nucleon interactions have been proposed and their exploration is now undergoing a very active and productive phase. Phenomenologically attractive extensions of the standard paradigm involve velocity and momentum dependent interactions~\cite{Chang:2009yt}, isospin violating couplings~\cite{Feng:2011vu}, and new long-range interactions~\cite{Fornengo:2011sz}. The study of these theoretical frameworks is still in progress. In this context, interesting results have for instance been found in studying anapole and magnetic dipole dark matter~\cite{DelNobile:2014eta}, light dark matter candidates~\cite{Gresham:2013mua}, dark matter capture by the Sun~\cite{Liang:2013dsa}, and benchmark models designed for dark matter searches at the LHC~\cite{deSimone:2014pda}. Momentum dependent interactions have also been explored in the context of extracting the phase-space distribution of dark matter particles with direct detection experiments~\cite{Peter:2013aha}. Halo-independent analysis based on velocity and momentum dependent operators can be found in Refs.~\cite{DelNobile:2013cva,Cherry:2014wia}.

Recently, Refs.~\cite{Fitzpatrick:2012ib,Fitzpatrick:2012ix,Anand:2013yka} proposed the idea of studying the dark matter-nucleon interaction with a non-relativistic effective theory approach similar to the one used in the 60's for exploring weak-interactions. Ref.~\cite{Fitzpatrick:2012ib} extends the work of Ref.~\cite{Fan:2010gt} to a systematic and complete classification of dark matter-nucleon non-relativistic interactions under Galilean transformations and conservation of energy and momentum. A general method to translate experimental limits into constraints on the dark matter-nucleon couplings in the non-relativistic effective theory has been devised~\cite{DelNobile:2013sia,Panci:2014gga}. Publicly available {\sffamily Mathematica} packages to perform these calculations are also available~\cite{Anand:2013yka,DelNobile:2013sia}.  Exclusion limits for velocity and momentum dependent interaction operators have been obtained from single experiments in~\cite{Fitzpatrick:2012ix,Gresham:2014vja,Buckley:2013jwa}. 

So far, the analysis of the non-relativistic effective theory has (1) considered the different interaction operators separately and (2) analyzed distinct direct detection experiments independently. A global analysis of the full multidimensional parameter space defining the effective theory of the dark matter-nucleon interaction is still missing. To tackle this challenge is the main aim of this work. Here we present the first comprehensive analysis of the multidimensional parameter space of the dark matter-nucleon effective theory. In this study all the couplings and the dark matter mass are simultaneously considered as free parameters. In addition, we have combined in a single analysis many different direct detection data, including the recent LUX, SuperCDMS and CDMSlite results. To achieve these goals we have exploited state-of-the-art Bayesian/frequentist numerical tools to sample the posterior probability density function and the profile likelihood of the model parameters. Importantly, we find that present direct detection data contain sufficient information to simultaneously constrain all the interaction operators present in the effective theory of the dark matter-nucleon interaction.

The paper is organized as follows. In Sec.~\ref{theory} we briefly review the non-relativistic effective theory of the dark matter direct detection proposed in Ref.~\cite{Fitzpatrick:2012ib}. Sec.~\ref{statistics} describes the statistical methods used in our analysis, whereas the datasets to which they are applied are introduced in Sec.~\ref{datasets}. Secs.~\ref{result1} and~\ref{result2} are devoted to the presentation of the results, and Sec.~\ref{conclusions} contains our conclusions. Appendix~\ref{astromcmc} describes the dependence of our results on the astrophysical assumptions, whereas Appendix~\ref{dmrfun} contains a list of the dark matter response functions relevant for this paper.

\section{Effective theory of the dark matter-nucleon interaction}
\label{theory}
In this section we review the basic concepts and equations defining the effective theory of the dark matter-nucleon interaction. For a more detailed introduction to this subject we refer the reader to the original literature~\cite{Fan:2010gt,Fitzpatrick:2012ib,Fitzpatrick:2012ix,Anand:2013yka}.  

From the point of view of relativistic quantum field theory, effective dark matter-nucleon interactions can be constructed from Lorentz-invariant combinations of dark matter and nucleonic bilinear operators. In the dark matter-nucleon non-relativistic effective theory the interactions are restricted by Galilean invariance, energy and momentum conservation, and hermiticity~\cite{Fitzpatrick:2012ib}. These requirements allow to construct a generating set of five non-relativistic operators for the algebra of $\chi$-nucleon effective interaction operators (here $\chi$ denotes the dark matter particle): the identity $1_\chi 1_N$, the momentum transfer\footnote{Our definition of momentum transfer $\vec{q}$ is the common one in dark matter direct detection studies, namely $\vec{q} = \vec{p}_{\chi}-\vec{p}^{\,\prime}_{\chi}$, where $\vec{p}_{\chi}$ and $\vec{p}^{\,\prime}_{\chi}$ are the initial and final dark matter momenta. Ref.~\cite{Fitzpatrick:2012ib} defines $\vec{q}$ with opposite sign. This explains the minus signs in the $\vec{q}$-dependent operators in Tab.~1 and in the expression for $\vec{v}^{\perp}_{\chi N}$.} $\vec{q}$, the $\chi$-nucleon transverse relative velocity operator $\vec{v}^{\perp}_{\chi N}$ (with matrix element equal to $\vec{v}_{\chi N}-\vec{q}/2\mu_N$, where $\vec{v}_{\chi N}$ is the initial $\chi$-nucleon relative velocity and $\mu_N$ is the $\chi$-nucleon reduced mass), and the dark matter and nucleon spin operators $\vec{S}_\chi 1_N$ and $1_\chi \vec{S}_{N}$, respectively.  The most general effective theory at the dark matter-nucleon level involves products of the five generating operators. In this paper, we restrict ourselves to the exchange of a heavy spin-0 or spin-1 particle, and following Ref.~\cite{Fitzpatrick:2012ib}, we limit ourselves to the 10 operators listed in Tab.~1. The additional operator $\mathcal{O}_2=(v^\perp_{\chi N})^2$ cannot be a leading-order operator in effective theories, and the remaining operators $\mathcal{O}_{16}=-\mathcal{O}_{10}\mathcal{O}_{5}$,   $\mathcal{O}_{13}=\mathcal{O}_{10}\mathcal{O}_{8}$,  $\mathcal{O}_{15}=-\mathcal{O}_{11}\mathcal{O}_{3}$ and $\mathcal{O}_{14}=\mathcal{O}_{11}\mathcal{O}_{7}$ are difficult to generate in explicit particle models. For spin-0 dark matter particles, the spin operator $\vec{S}_{\chi}$ is identically zero. For spin-1/2 particles it is equal to $\vec{\sigma}/2$, where $\sigma_i$, $i=1,2,3$, are the Pauli sigma matrices acting on the $\chi$-spinor. For spin-1 dark matter particles the components of $\vec{S}_{\chi}$ are spin-1 representations of the angular momentum generators acting on the $\chi$-vector. 

\begin{table}
    \centering
    \begin{tabular}{ll}
    \toprule
        $\mathcal{O}_1 = 1_{\chi} 1_{N}$ \hspace{10em} &         $\mathcal{O}_7 = \vec{S}_{N}\cdot \vec{v}^{\perp}_{\chi N}$ \\
        $\mathcal{O}_3 = -i\vec{S}_N\cdot\left(\frac{\vec{q}}{m_N}\times\vec{v}^{\perp}_{\chi N}\right)$ &         $\mathcal{O}_8 = \vec{S}_{\chi}\cdot \vec{v}^{\perp}_{\chi N}$ \\
        $\mathcal{O}_4 = \vec{S}_{\chi}\cdot \vec{S}_{N}$ &         $\mathcal{O}_9 = -i\vec{S}_\chi\cdot\left(\vec{S}_N\times\frac{\vec{q}}{m_N}\right)$ \\                                                                             
        $\mathcal{O}_5 = -i\vec{S}_\chi\cdot\left(\frac{\vec{q}}{m_N}\times\vec{v}^{\perp}_{\chi N}\right)$ &         $\mathcal{O}_{10} = -i\vec{S}_N\cdot\frac{\vec{q}}{m_N}$ \\                                                                                                        
        $\mathcal{O}_6 = \left(\vec{S}_\chi\cdot\frac{\vec{q}}{m_N}\right) \left(\vec{S}_N\cdot\frac{\vec{q}}{m_N}\right)$ &        $\mathcal{O}_{11} = -i\vec{S}_\chi\cdot\frac{\vec{q}}{m_N}$ \\                                                                                                      
    \bottomrule
    \end{tabular}
    \caption{List of the 10 non-relativistic operators defining the effective theory of the dark matter-nucleon interaction studied in this paper. The operators $\mathcal{O}_i$ are the same as in Ref.~\cite{Anand:2013yka}.}
    \label{operators}
\end{table}

The most general effective Hamiltonian describing the dark matter interaction with a point-like nucleon is then given by the following linear combination of operators 
\begin{equation}
\mathcal{H} = \sum_i \left( c^{0}_i + c^{1}_i \tau_3 \right) \mathcal{O}_i \,.
\label{Hamiltonian}
\end{equation}
Here $\tau_3$ is the third isospin Pauli matrix.
The coupling constants $c^{\tau}_i$ ($\tau=0,1$) have dimension (mass)$^{-2}$, and are analogous to the Fermi constant $G_F$.\footnote{We define the $c^\tau_i$ constants following Ref.~\cite{Anand:2013yka}. Other definitions exist in the literature. For example, Ref.~\cite{DelNobile:2013sia} has $\mathfrak{c}^{\tau}_1 = 4 m_\chi m_N c^{\tau}_1$.}
The constants $c_i^0$ correspond to isoscalar dark matter-nucleon interactions, whereas the constants $c_i^1$ describe the isovector interactions. Equivalently, $c^{\rm p}_i = (c^{0}_i+c^{1}_i)/2$ and $c^{\rm n}_i = (c^{0}_i-c^{1}_i)/2$ are the coupling constants for protons and neutrons, respectively. In this paper we restrict our analysis to isoscalar interactions (often but improperly called ``isospin-conserving'' interactions), i.e., we set $c^{1}_i=0$ (see Ref.~\cite{Cirigliano:2013zta} for an analysis of isovector couplings). The interaction Hamiltonian used to calculate the cross section for dark matter scattering on nucleons bound in a detector nucleus is obtained from Eq.~(\ref{Hamiltonian}) by replacing the point-like charge and spin operators with the corresponding extended nuclear charge and spin-current densities, as for instance in Eq.~27 of Ref.~\cite{Anand:2013yka}. In this case the relative $\chi$-nucleon transverse velocity operator $\vec{v}^{\perp}_{\chi N}$ is conveniently rewritten as $\vec{v}^{\perp}_{\chi N}=\vec{v}^{\perp}_{\chi T}-\vec{v}^{\perp}_{NT}$~\cite{Fitzpatrick:2012ib}, where the first term $\vec{v}^{\perp}_{\chi T}$ is the $\chi$-nucleus transverse velocity operator (with matrix element equal to $\vec{v}_{\chi T}-\vec{q}/2\mu_T$, where $\vec{v}_{\chi T}$ is the initial $\chi$-nucleus relative velocity and $\mu_T$ is the $\chi$-nucleus reduced mass), and the second term $\vec{v}^{\perp}_{NT}$ is the transverse relative velocity of the nucleon $N$ with respect to the nucleus center of mass~\cite{Fitzpatrick:2012ib}. To simplify the notation and connect it to the usual notation in analyses of dark matter experiments, we write $\vec{v}$ without index for the relative $\chi$-nucleus velocity $\vec{v}_{\chi T}$. 

The differential cross section for dark matter scattering on a target nucleus of mass $m_T$ is given by
\begin{equation}
\frac{d\sigma}{dE_{R}} = \frac{m_{T}}{2\pi v^2} \Bigg[ \frac{1}{2j_\chi+1}\frac{1}{2j_N+1} \sum_{\rm spins} |\mathcal{M}_{NR}|^2 \Bigg]
\label{dsigmadER}
\end{equation}
where $|\mathcal{M}_{NR}|^2$ denotes the square modulus of the non-relativistic scattering amplitude $\mathcal{M}_{NR}$ (related to the usual invariant amplitude $\mathcal{M}$ by $\mathcal{M} = 4 m_T^2 \mathcal{M}_{NR}$), and $j_\chi$ and $j_N$ are the dark matter and nucleus spins, respectively. When averaged over initial spins and summed over final spins, $|\mathcal{M}_{NR}|^2$ gives a quantity $P_{\rm tot}$ proportional to the total transition probability, which can be expressed as a combination of nuclear and dark matter response functions. In the most general case it takes the following form 
\allowdisplaybreaks
\begin{eqnarray}
P_{\rm tot}({v}^2,{q}^2)&\equiv&{1 \over 2j_\chi + 1} {1 \over 2j_N + 1} \sum_{\rm spins} |\mathcal{M}_{NR}|^2 \nonumber \\  &=& {4 \pi \over 2j_N + 1} 
\sum_{ \tau=0,1} \sum_{\tau^\prime = 0,1} \Bigg\{ \Bigg[ R_{M}^{\tau \tau^\prime}({v}^{\perp 2}_{\chi T}, {{q}^{2} \over m_N^2})~W_{M}^{\tau \tau^\prime}(y)   \nonumber\\
&+& R_{\Sigma^{\prime \prime}}^{\tau \tau^\prime}({v}^{\perp 2}_{\chi T}, {{q}^{2} \over m_N^2})   ~W_{\Sigma^{\prime \prime}}^{\tau \tau^\prime}(y) 
+   R_{\Sigma^\prime}^{\tau \tau^\prime}({v}^{\perp 2}_{\chi T}, {{q}^{2} \over m_N^2}) ~ W_{\Sigma^\prime}^{\tau \tau^\prime}(y) \Bigg]  \nonumber\\  
&+& {{q}^{2} \over m_N^2} ~\Bigg[R_{\Phi^{\prime \prime}}^{\tau \tau^\prime}({v}^{\perp 2}_{\chi T}, {{q}^{2} \over m_N^2}) ~ W_{\Phi^{\prime \prime}}^{\tau \tau^\prime}(y)  +  R_{ \Phi^{\prime \prime}M}^{\tau \tau^\prime}({v}^{\perp 2}_{\chi T}, {{q}^{2} \over m_N^2})  ~W_{ \Phi^{\prime \prime}M}^{\tau \tau^\prime}(y) \nonumber\\
&+&   R_{\tilde{\Phi}^\prime}^{\tau \tau^\prime}({v}^{\perp 2}_{\chi T}, {{q}^{2} \over m_N^2}) ~W_{\tilde{\Phi}^\prime}^{\tau \tau^\prime}(y) 
+   R_{\Delta}^{\tau \tau^\prime}({v}^{\perp 2}_{\chi T}, {{q}^{2} \over m_N^2}) ~ W_{\Delta}^{\tau \tau^\prime}(y) \nonumber\\
 &+&  R_{\Delta \Sigma^\prime}^{\tau \tau^\prime}({v}^{\perp 2}_{\chi T}, {{q}^{2} \over m_N^2})  ~W_{\Delta \Sigma^\prime}^{\tau \tau^\prime}(y)   \Bigg]  \Bigg\}  \,.
\label{Ptot}
\end{eqnarray}
Notice that
\begin{align}
v^{\perp 2}_{\chi T} = v^2 - \frac{q^2}{4\mu_T^2} .
\end{align}
For completeness, we list the dark matter response functions $R_{M}^{\tau \tau^\prime}$, $R_{\Sigma^{\prime \prime}}^{\tau \tau^\prime}$, $R_{\Sigma^\prime}^{\tau \tau^\prime}$, $R_{\Phi^{\prime \prime}}^{\tau \tau^\prime}$, $R_{\Phi^{\prime\prime}M}^{\tau \tau^\prime}$, $R_{\tilde{\Phi}^\prime}^{\tau \tau^\prime}$, $R_{\Delta}^{\tau \tau^\prime}$ and $R_{\Delta \Sigma^\prime}^{\tau \tau^\prime}$ in Appendix~\ref{dmrfun}. The nuclear response functions  $W_{M}^{\tau \tau^\prime}$, $W_{\Sigma^{\prime \prime}}^{\tau \tau^\prime}$, $W_{\Sigma^\prime}^{\tau \tau^\prime}$, $W_{\Phi^{\prime \prime}}^{\tau \tau^\prime}$, $W_{\Phi^{\prime\prime}M}^{\tau \tau^\prime}$, $W_{\tilde{\Phi}^\prime}^{\tau \tau^\prime}$, $W_{\Delta}^{\tau \tau^\prime}$ and $W_{\Delta \Sigma^\prime}^{\tau \tau^\prime}$ can be evaluated for different target materials and isotopes using the {\sffamily Mathematica} package of Ref.~\cite{Anand:2013yka}, or the approximate expressions provided in the appendix of Ref.~\cite{Fitzpatrick:2012ib}, where the functions $F_{IJ}^{\tau\tau'} = 4\pi W_{IJ}^{\tau\tau'}/(2j_N+1)$. The definition of these nuclear response functions is given in Eq.~(41) of Ref.~\cite{Anand:2013yka}. In our calculations we have rewritten the {\sffamily Mathematica} package of Ref.~\cite{Anand:2013yka} in {\sffamily FORTRAN}, and used our own routines to calculate differential cross sections and scattering rates. In Eq.~(\ref{Ptot}), $y=(q b/2)^2$, where $b$ is the   oscillator parameter in the independent-particle harmonic oscillator model ~\cite{Anand:2013yka}.

The differential rate of scattering events per unit time and per unit detector mass is obtained as
\begin{equation}
\frac{d\mathcal{R}}{dE_{R}} =  \sum_{T}\frac{d\mathcal{R}_{T}}{dE_{R}} \equiv  \sum_{T} \xi_T \frac{\rho_{\chi}}{2\pi m_\chi}  \left\langle  \frac{1}{v} P_{\rm tot}(v^2,q^2)  \right\rangle
\label{rate_theory}
\end{equation}
where $\xi_T$ is the mass fraction of the nucleus $T$ in the target material, $\rho_\chi$ is the local dark matter density, and $m_\chi$ is the dark matter mass. The angle brackets in Eq.~(\ref{rate_theory}) denote an average over the local dark matter velocity distribution, $f$, in the galactic rest frame boosted to the detector frame, namely 
\begin{equation}
\left\langle \frac{1}{v} P_{\rm tot}(v^2,q^2)  \right\rangle =  \int\limits_{v>v_{\rm min}(q)} \,  \frac{f(\vec{v} + \vec{v}_e(t))}{v} \, P_{\rm tot}(v^2,q^2) \, d^3v  , 
\end{equation} 
where $\vec{v}_e(t)$ is the time-dependent Earth velocity in the galactic rest frame, and $v_{\rm min}(q)=q/2\mu_T$ is the minimum velocity required for a dark matter particle to transfer a momentum $q$ to the target nucleus. In our calculations we consider two choices of $f$: a Maxwell-Boltzmann distribution $f(\vec{v} + \vec{v}_e(t))\propto \exp(-|\vec{v}+\vec{v}_{e}(t)|^2/v_0^2)$ truncated at the local escape velocity $v_{\rm esc}$ , and the anisotropic velocity distribution proposed in Ref.~\cite{Bozorgnia:2013pua}.

\section{Statistical framework}
\label{statistics}
In this section we introduce the statistical methods used to extract limits on the strength of the dark matter-nucleon effective interactions from present dark matter direct detection data.  We present both a Bayesian approach~\cite{Trotta:2006ew,Akrami:2010dn,Akrami:2010cz,Pato:2010zk,Arina:2011si,Arina:2011zh,Peter:2011eu,Arina:2012dr,Strege:2012kv,Arina:2013jma,Peter:2013aha,Cerdeno:2013gqa,Cerdeno:2014uga} and a frequentist approach. This allows a comprehensive analysis of the multidimensional parameter space studied in this paper. See Ref.~\cite{Cowan} for an introduction to Bayesian and frequentist statistical methods.

In the Bayesian analysis, our efforts are concentrated on reconstructing the posterior probability density function (PDF) of the model parameters, $\mathcal{P}(\mathbf{\Theta}|\mathbf{d})$. The posterior PDF depends on the $n$ datasets $\mathbf{d}=(d_1,\dots,d_n)$ considered in the analysis and on an array of $m$ free parameters, denoted by $\mathbf{\Theta}=(\theta_1,\cdots,\theta_m)$. The posterior PDF describes our degree of belief in a certain hypothesis after having considered the data (e.g., the validity of a specific configuration in parameter space). It is related to the likelihood function $\mathcal{L}(\mathbf{d}|\mathbf{\Theta})$ by Bayes' theorem, \begin{equation}
\mathcal{P}(\mathbf{\Theta}|\mathbf{d}) = \frac{\mathcal{L}(\mathbf{d}|\mathbf{\Theta}) \pi(\mathbf{\Theta})}{\mathcal{E}(\mathbf{d})}\,.
\end{equation} 
In this expression $\pi(\mathbf{\Theta})$ is the prior PDF, which describes our degree of belief in a certain hypothesis {\it before} having seen the available data. The Bayesian evidence $\mathcal{E}(\mathbf{d})$ is an important concept in the model comparison. Being independent of the model parameters, however, it simply plays the role of a normalization constant when performing parameter inference, as in the present analysis. 

The parameter space explored in our investigations  is spanned by the coupling constants $c^{\tau}_{i}$, with $\tau=0,1$ and $i=1,3,\dots,11$, the dark matter mass, $m_{\chi}$, and a set of ``nuisance'' parameters $\bfeta$ introduced to model different sources of uncertainties affecting the interpretation of the data. There are two types of nuisance parameters relevant for the present analysis. A first type concerns the local dark matter space and velocity distribution. In our investigations - which focus on the form of the dark matter-nucleon interaction - we set the astrophysical nuisance parameters in a configuration known as the ``standard dark matter halo''~\cite{2012arXiv1209.3339F}. This is characterized by a truncated Maxwell-Boltzmann dark matter velocity distribution (in the galactic rest frame), with $v_0=220$~km~s$^{-1}$, escape velocity $v_{\rm esc}=544$~km~s$^{-1}$ and a local dark matter density $\rho_{\chi}=0.3$~GeV~cm$^{-3}$. In two examples, presented in appendix~\ref{astromcmc}, we relax this assumption, considering a more general astrophysical setup characterized by 8 parameters describing a galactic dark matter component with an anisotropic velocity distribution~\cite{Bozorgnia:2013pua}. The second type of nuisance parameters introduced in our statistical analysis is instead related to the presence of poorly known experimental quantities affecting the calculation of the expected dark matter direct detection signals  (e.g., quenching factors, threshold effects, etc.). These parameters are introduced in the next section, describing the datasets included in this study. They are always treated as free parameters. Tab.~\ref{prior} summarizes the free parameters considered in the following analysis. Following~\cite{Anand:2013yka}, we have introduced the mass scale $m_v=246.2$ GeV and the dimensionless quantities $c^{\tau}_i m_v^2$. 

Among the many interesting pieces of information that can be obtained from the knowledge of the posterior PDF, we concentrate on 1D and 2D marginal posterior PDFs. These are calculated integrating the posterior PDF over the other model parameters. For instance, the 2D marginal posterior PDF of the parameters $\theta_1$ and $\theta_2$ (e.g., $c^0_1$ and $m_{\chi}$) can be obtained integrating, i.e., marginalizing, over the remaining parameters as follows
\begin{equation}
\mathcal{P}_{\rm marg}(\theta_1,\theta_2|\mathbf{d})  \propto \int d\theta_3\dots d\theta_m\, \mathcal{P}(\mathbf{\Theta}|\mathbf{d}) \,.
\label{marg}
\end{equation} 
Limits on the coupling constants $c_i^{\tau}$ are then expressed in terms of $x$\% credible regions, defined as the portions of the parameter space containing $x$\% of the total posterior probability and such that $\mathcal{P}_{\rm marg}$ at any point inside the region is larger than at any point outside the region. 

When the likelihood function is well approximated by a multivariate Gaussian and contains more information than the prior PDF, the associated credible regions tend to favor the portion of parameter space where the likelihood function is near its maximum $\mathcal{L}_{\rm max}$. However, this is not true in general. For datasets containing an insufficient amount of information like those studied here,  the integral in Eq.~(\ref{marg}) may be dominated by the tails of the posterior PDF, if these tails extend over a large volume of parameter space. A useful statistical indicator that is insensitive to these ``volume effects'' is the D-dimensional profile likelihood, which in the 2D case is defined as follows~\cite{2011JHEP...06..042F} 
\begin{equation}
\mathcal{L}_{\rm prof}(\mathbf{d}|\theta_1,\theta_2) \propto \max_{\theta_3,\dots,\theta_m} \mathcal{L}(\mathbf{d}|\mathbf{\Theta}) \,.
\label{eq:prof_likelihood}
\end{equation} 
While the profile likelihood does not admit a formal interpretation in terms of a probability density function, it can conventionally be used to construct approximate frequentist confidence intervals from an effective chi-square defined as $\Delta \chi^2_{\rm eff}\equiv-2 \ln \mathcal{L}_{\rm prof}/ \mathcal{L}_{\rm max}$. Wilks' theorem guarantees that under certain regularity conditions the distribution of $\Delta \chi^2_{\rm eff}$ converges to a chi-square distribution with, e.g., 2 degrees of freedom in the case of a 2D profile likelihood~\cite{2011JHEP...06..042F}.

\begin{table}
    \centering
    \begin{tabular}{lclcc}
    \toprule
    Parameter         & Type & Prior range &  Prior type & Reference \\
    \midrule                                          
    $\log_{10} (c_1^\tau m_v^2)$ & model parameter & $[-5,1]$ & log-prior & - \\
    $\log_{10} (c_3^\tau m_v^2)$ & model parameter & $[-4,4]$ & log-prior & - \\
    $\log_{10} (c_4^\tau m_v^2)$ & model parameter & $[-2,3]$ & log-prior & - \\
    $\log_{10} (c_5^\tau m_v^2)$ & model parameter & $[-4,4]$ & log-prior & - \\
    $\log_{10} (c_6^\tau m_v^2)$ & model parameter & $[-4,4]$ & log-prior & - \\
    $\log_{10} (c_7^\tau m_v^2)$ & model parameter & $[-4,4]$ & log-prior & - \\
    $\log_{10} (c_8^\tau m_v^2)$ & model parameter & $[-4,4]$ & log-prior & - \\
    $\log_{10} (c_9^\tau m_v^2)$ & model parameter & $[-4,4]$ & log-prior & - \\
    $\log_{10} (c_{10}^\tau m_v^2)$ & model parameter & $[-4,4]$ & log-prior & - \\
    $\log_{10} (c_{11}^\tau m_v^2)$ & model parameter & $[-4,4]$ & log-prior & - \\
    $\log_{10} (m_{\chi}/{\rm GeV})$ & model parameter & $[0.5 (0.1),4]$  & log-prior & - \\
     \midrule  
     $q_{\rm Na}$ & nuisance &   $[0.2,0.4]$ & Gaussian. $\sigma=0.10$ & Ref.~\cite{Schwetz:2011xm} \\		
     $\xi_{\rm Xe}$ & nuisance &   $[0.78,0.86]$ & Gaussian. $\sigma=0.04$ & Ref.~\cite{Arina:2011si} \\	
     $a_{\rm COUPP}$ & nuisance &   $[0.13,0.17]$ & Gaussian. $\sigma=0.02$ & Ref.~\cite{Behnke:2012ys} \\	
     $a_{\rm PICASSO}$ & nuisance &   $[2.5,7.5]$ & Gaussian. $\sigma=2.50$ & Ref.~\cite{Archambault:2012pm} \\	
     $a_{\rm SIMPLE}$ & nuisance &   $[3.34,3.86]$ & Gaussian. $\sigma=0.26$ & Ref.~\cite{Felizardo:2011uw} \\	
     $t_{\rm max}$ [days] & nuisance &   $[58,154]$ & Gaussian. $\sigma=24$ & Ref.~\cite{Aalseth:2014jpa} \\			 
    \bottomrule
    \end{tabular}
    \caption{List of model parameters and nuisance parameters. Together with the type of prior, we also report the prior range and the reference from which this range has been taken. We have chosen the prior ranges for $c^0_{1}$ and $c^0_{4}$ in light of the already existing experimental limits on these coupling constants. Gaussian prior PDFs are characterized by a mean lying at the center of the prior range and a standard deviation given by $\sigma$. The nuisance parameters $q_{\rm Na}$, $\xi_{\rm Xe}$, $a_{\rm COUPP}$, $a_{\rm PICASSO}$ and $a_{\rm SIMPLE}$ are introduced in Sec.~\ref{datasets} to model various types of detector uncertainties, whereas $t_{\rm max}$ is the peak date used to describe the CoGeNT modulation signal~\cite{Aalseth:2014jpa}. Regarding the lower bound of the dark matter mass prior range we have considered 0.1 for PICASSO, SuperCDMS and CDMSlite, and 0.5 for the other experiments. Following~\cite{Anand:2013yka}, we have expressed the coupling constants in units of $m_v^{-2}=(246.2~{\rm GeV})^{-2}$. }
    \label{prior}
\end{table}
Within this approach to data analysis, all the experimental information is encoded in the likelihood function, and, to some extent, in the choice of the prior PDF (when calculating the posterior PDF), if specific assumptions are made in order to give more weight to certain portions of the parameter space. If not otherwise specified, in the analysis we use a Poisson likelihood to model the distribution of the observed data. This is an appropriate choice when the datasets consist of a small sample of $k$ events, as for the recoil events detected (or searched for) by current dark matter direct detection experiments. Therefore, neglecting an irrelevant (for the parameter inference) constant term, our ``default'' choice for the likelihood function is~\cite{Akrami:2010dn,Akrami:2010cz,Pato:2010zk,Arina:2011si,Arina:2011zh,Peter:2011eu,Arina:2012dr,Strege:2012kv,Arina:2013jma,Peter:2013aha,Cerdeno:2013gqa,Cerdeno:2014uga}
\begin{equation}
-\ln \mathcal{L}(\mathbf{d}|m_\chi,\mathbf{c},\bfeta,\mu_B) = \mu_S(m_\chi,\mathbf{c},\bfeta)+\mu_B - k \ln [\mu_S(m_\chi,\mathbf{c},\bfeta)+\mu_B] \,,
\label{Like}
\end{equation}    
where $\mu_S(m_\chi,\mathbf{c},\bfeta)$ represents the expected number of scattering events. For every experiment, we calculate $\mu_S(m_\chi,\mathbf{c},\bfeta)$ within the effective theory of the dark matter-nucleon interaction. $\mu_S(m_\chi,\mathbf{c},\bfeta)$ depends on the dark matter mass $m_\chi$, the coupling constants $\mathbf{c}=(c^0_1,c^1_1,\dots,c^0_{11},c^1_{11})$ and a set of nuisance parameters $\bfeta$ characteristic of the experiment under analysis. The likelihood in Eq.~(\ref{Like}) also depends on the expected (or measured) number of background events $\mu_B$. For some of the experiments considered here, this background is a stochastic variable with a Gaussian distribution of variance $\sigma_{B}^2$ and average $\hat{\mu}_B$. In this case, one can marginalize over the experimental background analytically, obtaining in the limit $\sigma_B \ll \hat{\mu}_B$ the form of the effective likelihood actually implemented as default choice in our analysis, namely  
\begin{eqnarray}
-\ln \mathcal{L}_{\rm eff}(\mathbf{d}|m_\chi,\mathbf{c},\bfeta) &=& -\ln \left\{  \int d\mu_B \, \frac{e^{-\frac{\left(\mu_B-\hat{\mu}_B\right)^2}{2\sigma_B^2}} }{\sqrt{2\pi\sigma_B^2}}
\frac{[\mu_S(m_\chi,\mathbf{c},\bfeta)+\mu_B]^k}{k!} e^{-[\mu_S(m_\chi,\mathbf{c},\bfeta)+\mu_B]} \right\} \nonumber\\
&&\nonumber \\
&\simeq& \mu_S(m_\chi,\mathbf{c},\bfeta)+\hat{\mu}_B + (2-k) \ln [\mu_S(m_\chi,\mathbf{c},\bfeta)+\hat{\mu}_B]  \nonumber\\
&-&\ln \left\{ \frac{(k^2-k)}{2}\sigma_B^2 + \left[ \mu_S(m_\chi,\mathbf{c},\bfeta)+\hat{\mu}_B - \frac{k}{2}\sigma_B^2 \right]^2 \right\} \,.
\label{Like_eff}
\end{eqnarray}    
In the second line we have expanded the Poisson factor around $\mu_B=\hat{\mu}_B$ and kept only the leading terms in this expansion. This procedure is justified by the fact that in the limit $\sigma_B \ll \hat{\mu}_B$ (a reasonable assumption for all the experiments that we consider in the paper), the Gaussian factor in the integrand of Eq.~(\ref{Like_eff}) tends to the Dirac delta $\delta(\mu_B-\hat{\mu}_B)$. 

Within our investigations, we employ log-priors both for the dark matter mass and for the coupling constants $\mathbf{c}$, namely 
\begin{equation}
\pi(\mathbf{\Theta}) \propto \prod_{i=1}^{m} \left[ \Theta_{\rm H}(\ln \theta_i-\ln\theta_i^{\rm min})-\Theta_{\rm H}(\ln\theta_i-\ln\theta_i^{\rm max})\right]
\label{eq:prior}
\end{equation}
where $\Theta_{\rm H}$ is the Heaviside theta-function and $\theta_i^{\rm min}$ and $\theta_i^{\rm max}$ are the extrema of the prior ranges shown in Tab.~\ref{prior}. This assumption allows to sample the posterior PDF varying the model parameters within prior ranges spanning several orders of magnitude.  

To sample the multidimensional likelihood surface and therefore reconstruct the posterior PDF and the profile likelihood of the model parameters, we use the {\sffamily Multinest} program~\cite{Feroz:2008xx,Feroz:2007kg,Feroz:2013hea}. We use our own routines to calculate the scattering rates predicted by the dark matter-nucleon effective theory and to evaluate the likelihood function. Figures have been produced using the programs {\sffamily GetDist}~\cite{Lewis:2002ah},  {\sffamily Getplots}~\cite{Austri:2006pe} and {\sffamily Matlab}. When calculating the profile likelihood we set the {\sffamily Multinest} parameters to $n_{\rm live}=20000$ and ${\rm tol}=10^{-4}$, producing approximately $3\times10^{6}$ likelihood evaluations.   

\section{Datasets and likelihoods}
\label{datasets}
We now introduce the data used in our analysis of the dark matter-nucleon effective interactions, providing the details required in order to evaluate Eq.~(\ref{Like_eff}). To this aim, we first notice that in a real experiment the theoretical rate in Eq.~(\ref{rate_theory}) is not the quantity directly observed. In many cases the measurable energy $E_{\mathcal{O}}$ is only a fraction of the true nuclear recoil energy $E_{R}$ deposited by a dark matter particle in the detector. Scintillators are an important example of detectors with this property. Moreover, the finite energy resolution and the limited efficiency $\mathcal{E}$ of a real detector can affect the observed direct detection rates.  For a Gaussian energy resolution function (see Eqs.~(\ref{xenon100a})--(\ref{xenon100}) for a non-Gaussian example), we write the observable differential rate of scattering events per unit time and per unit detector mass as follows 
\begin{equation}
\label{rate_obs}
\frac{d\mathcal{R}}{d\hat{E}_{\mathcal{O}}}=  \mathcal{E}(\hat{E}_{\mathcal{O}}) \int_{0}^{\infty} dE_{\mathcal{O}} (2\pi\sigma^2)^{-1/2} \exp\left[-\frac{(E_{\mathcal{O}}-\hat{E}_{\mathcal{O}})^2}{2\sigma^2}\right]\frac{\partial E_{R}}{\partial E_{\mathcal{O}}} \times \left( \frac{d\mathcal{R}}{dE_{R}}\right)_{E_{R}=E_{R}(E_{\mathcal{O}})} \,.
\end{equation}   
In this expression $\hat{E}_{\mathcal{O}}$ is the actually observed energy, whereas $E_{\mathcal{O}}$ is the energy potentially measurable. The latter coincides with the former only in the limit of infinite experimental resolution. The energy dispersion $\sigma$ is in general an energy dependent quantity. When not otherwise specified, we consider the time average of Eq.~(\ref{rate_obs}). The total number of events $\mu_S(m_\chi,\mathbf{c},\bfeta)$ in the signal region $[\hat{E}_1,\hat{E}_2]$ is then calculated integrating Eq.~(\ref{rate_obs}) over this energy range and multiplying the result by the experimental exposure $MT$ (in kg-days, e.g.),
\begin{align}
\mu_S(m_\chi,\mathbf{c},\bfeta) = MT\int_{\hat{E}_1}^{\hat{E}_2} \, \frac{d\mathcal{R}}{d\hat{E}_{\mathcal{O}}} \, d\hat{E}_{\mathcal{O}} .
\label{eq:muS}
\end{align}

\subsection{CDMS-Ge}
\label{CDMS-Ge}
The Cryogenic Dark Matter Search (CDMS II) experiment employs 19 germanium and 11 silicon detectors operating at cryogenic temperatures to search for dark matter through the observation of phonons and ionization. The final exposure of the experiment, in an analysis focused on a subset of the operating germanium detectors, is of 612 kg-days, corresponding to four periods of stable data taking between July 2007 and September 2008~\cite{Ahmed:2009zw}. In this period, the CDMS collaboration has observed two events in the acceptance region 10 - 100 keV at recoil energies 12.3 keV and 15.5 keV,  with an expected number of background surface events equal to 0.9~$\pm$~0.3. We include this information in our analysis using the default likelihood (\ref{Like_eff}), with $k=2$, $\hat{\mu}_{B}=0.9$ and $\sigma_{B}=0.3$. To evaluate this expression we assume a maximum experimental efficiency of 32\% at 20 keV, linearly decreasing towards lower and higher energies, reaching the value of 20\% at 10 keV and at 100 keV. The energy resolution adopted in the calculations features a dispersion $\sigma=0.2$. For the CDMS experimental apparatus $E_{\mathcal{O}}=E_{R}$.

\subsection{CDMS Low Threshold}
\label{CDMS-LT}
The CDMS collaboration has also performed a low-threshold analysis of the data collected during six runs between October 2006 and September 2008. In their analysis the recoil energy threshold was lowered to 2 keV, while keeping the upper bound of the signal region at 100 keV~\cite{Ahmed:2010wy}. Below 10 keV the discrimination between nuclear and electron recoils degrades and leads to a higher expected number of background events. Within this analysis, one of the germanium detectors, namely the fifth detector in the first tower of the CDMS detector array (T1Z5), has observed  38 candidate events within the signal region (36 of which between 2 keV and 20 keV; see Fig.~2 in Ref.~\cite{Ahmed:2010wy}). The CDMS collaboration has identified three possible sources of background contamination, namely zero-charge events, surface events, and bulk events, which can explain 75\% of the observed candidate events~\cite{Ahmed:2010wy}. We include the results of this analysis in our investigations adding to the total likelihood a term of type (\ref{Like_eff}) with $k=36$ (considering the interval 2 - 20 keV as signal region, as in Ref.~\cite{Farina:2011pw}) $\hat{\mu}_{B}=36\times0.75$ and $\sigma_{B}=0$. Within our analysis we assume the efficiency shown in the inset of Fig.~1 in Ref.~\cite{Ahmed:2010wy}, an energy dependent energy resolution featuring $\sigma=\sqrt{0.293^2+(0.056 E_{\mathcal{O}})^2}$, an exposure for T1Z5 of 241/8 kg-days, and $E_{\mathcal{O}}=E_{R}$.

\subsection{SuperCDMS}
The SuperCDMS experiment is an upgrade of CDMS II which features new hardware devices interfaced with fifteen 0.6-kg cylindrical germanium crystals forming five towers containing three crystals each. The SuperCDMS collaboration has recently presented data recorded between October 2012 and June 2013 by a subset of 7 germanium detectors, corresponding to a total exposure of 577 kg-days~\cite{Agnese:2014aze}. This analysis has identified 11 dark matter candidate events passing the three levels of data-selection criteria introduced by the collaboration to discriminate candidate signal events from background events within the predefined signal region 1.6 - 10 keV. Tab.~1 of Ref.~\cite{Agnese:2014aze} provides details regarding the number of events recorded by the 7 germanium detectors, and the number of background events expected for each detector separately. We include these data in our analysis adding a contribution of the form (\ref{Like_eff}) to the total likelihood for each SuperCDMS detector, except for the detectors T5Z2 and T5Z3 for which the estimated number of background events seems to be significantly smaller than the number of actually observed candidate events (contrary to the other detectors). To evaluate these contributions to the likelihood function we set $\mu_B$ and $\sigma_B$ as in Tab.~1 of Ref.~\cite{Agnese:2014aze} and calculate $\mu_S(m_\chi,\mathbf{c},\bfeta)$ for each detector using Eq.~(\ref{rate_obs}) with $\sigma=0.3$ and the detector efficiency shown in Fig.~1 of Ref.~\cite{Agnese:2014aze}, assuming an average exposure per detector equal to 577/7 kg-days.

\subsection{CDMSlite}
In a previous analysis the SuperCDMS experiment has collected data during 10 live days of dark matter search, using a single iZIP detector operating in a different mode (compared to previous studies) that yielded significantly better sensitivity to dark matter candidates of mass less than 10 GeV. This new operating mode is called CDMS Low Ionization Threshold Experiment, or simply CDMSlite~\cite{Agnese:2013jaa}. The nuclear recoil energy threshold associated with this experimental configuration is of 170 eV$_{\rm ee}$, corresponding to 841 eV$_{\rm nr}$ as one can see solving the non-linear equation~\cite{Agnese:2013jaa}
\begin{equation}
E_{R} = E_{\mathcal{O}} \left(1+\frac{e V_b}{\varepsilon_\gamma}\right)\left[1+\frac{e V_b}{\varepsilon_\gamma}Y(E_{R})\right]^{-1}
\end{equation} 
which relates the true nuclear recoil energy $E_{R}$ (measured in keV$_{\rm nr}$) to the observable energy $E_{\mathcal{O}}$  (measured in keV$_{\rm ee}$). In this expression $e V_b=69$ eV and $\varepsilon_\gamma=3$ eV, whereas $Y(E_{R})$ is the ionization yield. The Lindhard model predicts for the latter~\cite{Agnese:2013jaa} 
\begin{equation}
Y(E_{R}) = \bar{k} \frac{g(\varepsilon)}{1+\bar{k} g(\varepsilon)}
\end{equation}
where $g(\varepsilon)= 3\varepsilon^{0.15} + 0.7\varepsilon^{0.6} + \varepsilon$, $\varepsilon=11.5 E_{R} Z^{-7/3}$ and $\bar{k}=0.157$ for a germanium target. No background subtraction was applied to the collected data. This analysis has found that the average rate of nuclear recoils in the CDMSlite detector is $5.2 \pm 1$ counts/keV$_{\rm ee}$/kg-day between 0.2 and 1 keV$_{\rm ee}$, and $2.9 \pm 0.3$ counts/keV$_{\rm ee}$/kg-day between 2 and 7 keV$_{\rm ee}$. To include this information in our analysis, we have first calculated the expected average count rates in the CDMSlite detector using Eq.~(\ref{rate_obs}), with the efficiency reported in the inset of Fig.~1 in Ref.~\cite{Agnese:2013jaa} and $\sigma\rightarrow 0$. Then, we have added a term to the total likelihood given by
\begin{eqnarray}
-\ln \mathcal{L}_{\rm CDMSlite} &=& \frac{1}{2}\Theta_{\rm H}(\mathcal{R}_{[0.2,1]}-5.2)(\mathcal{R}_{[0.2,1]}-5.2)^2 \nonumber\\
&+&  \frac{1}{2}\Theta_{\rm H}(\mathcal{R}_{[2,7]}-2.9)(\mathcal{R}_{[2,7]}-2.9)^2/0.3^2 
\end{eqnarray}
where $\mathcal{R}_{[0.2,1]}$ and $\mathcal{R}_{[2,7]}$ represent the average rates between 0.2 and 1 keV$_{\rm ee}$, and between 2 and 7 keV$_{\rm ee}$, respectively. 

\subsection{XENON100}
\label{XENON100}
The XENON100 experiment uses liquid xenon to search for dark matter through the detection of ionization and scintillation signals produced by dark matter interactions in the active volume of the detector. in Ref.~\cite{Aprile:2012nq} the XENON100 collaboration has presented data collected in 13 months during 2011 and 2012, with an effective exposure of 34$\times$224.6 kg-days. The ionization signal (S2) and the direct scintillation signal (S1) are both detected by arrays of  photomultipliers (PMTs), and measured in numbers of photoelectrons (PE). The expected number of S1 photoelectrons $\nu(E_{R})$ produced by a nuclear recoil of energy $E_{R}$ is given by $\nu(E_{R})=E_{R} L_{\rm eff}(E_{R}) L_y S_{\rm nr}/S_{\rm ee}$, where $L_y = 2.28 \pm 0.04$ PE/keV$_{\rm ee}$ is the light yield, $S_{\rm ee} = 0.58$ and $S_{\rm nr} = 0.95$ are the electric field scintillation quenching factors for electron and nuclear recoils, and finally, $L_{\rm eff}(E_{R})$ is the energy dependent scintillation efficiency. The actually produced number $n$ of photoelectrons for a given energy $E_{R}$ is subject to Poisson fluctuations around $\nu(E_{R})$, and to uncertainties related to the scintillation efficiency, which has not been measured at energies below 3 keV$_{\rm nr}$. We model these uncertainties introducing a nuisance parameter $\xi_{\rm Xe}$, first proposed in Ref.~\cite{Arina:2011si} to logarithmically extrapolate the scintillation efficiency toward low energies. This gives
\begin{equation}
L_{\rm eff}(E_{R})  = \left\{
\begin{array}{ll}
\bar{L}_{\rm eff}(E_{R}) \qquad  \qquad  \qquad  \qquad  \qquad  \qquad  \qquad  \,\quad \,\,\,{\rm for} \,\, E_{R}/{\rm keV}_{\rm nr} \ge 3 \nonumber\\
\nonumber\\
\max\{\xi_{\rm Xe}[\ln(E_{R}/{\rm keV}_{\rm nr}) -\ln3]+0.09, 0\} \qquad {\rm for} \,\,1 < E_{R}/{\rm keV}_{\rm nr} < 3
\end{array} \right.
\end{equation}
where $\bar{L}_{\rm eff}(E_{R})$ is the best-fit scintillation efficiency reported in Fig.~1 of Ref.~\cite{Aprile:2011hi}. Importantly, the observed number of photoelectrons S1 does not coincide with $n$ when the finite resolution of the detector photomultipliers is taken into account. The analysis of Ref.~\cite{Aprile:2012nq} uses S1 to reconstruct the recoil energy of the candidate dark matter signal events and the ratio S2/S1 to discriminate signal events from background events. In that analysis, XENON100 observed 2 candidate signal events in the pre-defined nuclear recoil energy range 6.6 - 30.5 keV$_{\rm nr}$ (corresponding to S1 in the range 3 - 30 PE). This observation is consistent with the background expectation of 1.0 $\pm$ 0.2 events. 

To include XENON100 in our analysis, we calculate the differential spectrum of the variable S1, first converting the recoil energy spectrum (\ref{rate_theory}) into the spectrum of the expected number of photoelectrons $n$, namely 
\begin{equation}
\frac{d\mathcal{R}}{dn} = \int_{0}^{\infty}dE_{R} \,{\rm Poiss}(n|\nu(E_{R})) \frac{d\mathcal{R}}{dE_{R}} 
\label{xenon100a}
\end{equation}
and then convolving the resulting expression with a Gaussian filter, to model the resolution of the detector photomultipliers:
\begin{equation}
\frac{d\mathcal{R}}{dS1} = \mathcal{E}(S1)\sum_{n=1}^{+\infty} {\rm Gauss}(S1 | n,\sqrt{n}\sigma_{\rm PMT}) \frac{d\mathcal{R}}{dn} \,.
\label{xenon100}
\end{equation}
Following Ref.~\cite{Aprile:2011hx}, in the first step we have used a Poisson PDF of mean $\nu(E_{R})$ to sample the number of actually produced photoelectrons $n$ associated with a given recoil energy $E_{R}$. The Gaussian filter employed in the second step has mean $n$ and variance $n\sigma^2_{\rm PMT}$, with $\sigma_{\rm PMT}=0.5$. The detector efficiency $\mathcal{E}(S_1)$ relevant for this analysis is shown in Fig.~2 of Ref.~\cite{Aprile:2011hi}. Integrating Eq.~(\ref{xenon100}) between 3 and 30 photoelectrons and multiplying the result by the experimental exposure, we finally obtain the expected number of signal events, $\mu_S(m_\chi,\mathbf{c},\bfeta)$. This prediction is then used to evaluate the likelihood (\ref{Like_eff}) with $k=2$, $\hat{\mu}_{B}=1$ and $\sigma_B=0.2$. 

\subsection{XENON10}
\label{XENON10}
In a second study the XENON collaboration reanalyzed the data from a 12.5 live day dark matter search, collected between August 23 and September 14 2006, using the S2 signal only to measure the nuclear recoil energy of the detected events~\cite{Angle:2011th}. The relation between the S2 variable (measured in PE) and the observed nuclear recoil energy is $S2=\mathcal{Q}_y(\hat{E}_{\mathcal{O}}) \hat{E}_{\mathcal{O}}$, where the function $\mathcal{Q}_y(\hat{E}_{\mathcal{O}})$ can be extracted from Fig.~1 of Ref.~\cite{Angle:2011th}. This type of analysis allows a very low recoil energy threshold (about 1.4 keV$_{\rm nr}$), increasing thus the detector sensitivity to low mass dark matter candidates, even if the detector ability in discriminating and rejecting electromagnetic background events is reduced within this setup. The experimental exposure corresponding to these data is 12.5$\times$1.2 kg-days. Within this analysis XENON10 observed 23 candidate events in the signal region 1.4 - 10 keV$_{\rm nr}$. XENON10 has also observed several dozens of single S2 electron events at lower energies, the origin of which is not clear yet. In our analysis we treat the 23 events in the signal region as possible dark matter candidates and estimate the expected nuclear recoil energy spectrum in XENON10 as (see Eq.~5.5 in Ref.~\cite{Lewin:1995rx})
\begin{equation}
\frac{d\mathcal{R}}{d\hat{E}_{\mathcal{O}}} = \mathcal{E}(\hat{E}_{\mathcal{O}})\left(\mathcal{Q}_y + \frac{\partial \mathcal{Q}_y}{\partial \hat{E}_{\mathcal{O}}}\hat{E}_{\mathcal{O}}\right) \int_{0}^{\infty}dE_{R} \,{\rm Poiss}(S2|\mathcal{Q}_y E_{R}) \frac{d\mathcal{R}}{dE_{R}}
\label{xenon10}
\end{equation}
assuming a constant efficiency $ \mathcal{E}(\hat{E}_{\mathcal{O}})=0.94$. In addition, we also include the possibility that a source of background events of unknown origin and characterized by a large error contributes to the observed events. The XENON10 contribution to the total likelihood is then estimated using Eq.~(\ref{Like_eff}) with $k=23$, $\hat{\mu}_B=20$ and $\sigma_B=10$. $\mu_S(m_\chi,\mathbf{c},\bfeta)$ is obtained integrating Eq.~(\ref{xenon10}) between 1.4 keV$_{\rm nr}$ and 10 keV$_{\rm nr}$, and multiplying the result by the experimental exposure.

\subsection{LUX}
\label{LUX}
The Large Underground Xenon (LUX) experiment is a dual-phase (liquid and gas) time-projection chamber with 250 kg of active volume. Similarly to XENON100, LUX searches for dark matter through the observation of prompt scintillation (S1) and ionization electrons, extracted into the gas portion of the detector, where they produce electroluminescence (S2)~\cite{Akerib:2013tjd}. In the case of LUX, the conversion between nuclear recoil energy (in keV$_{\rm nr}$) and number of photoelectrons can be extracted from the panel (b) of  Fig.~3 in Ref.~\cite{Akerib:2013tjd}.
The recent first data release consists of 85.3 live days of dark matter search data, collected between April 21 and August 8 2013. In this period, LUX observed 160 events with S1 between 2 and 30 PE, only one of which is (slightly) below the mean of the Gaussian fit to the nuclear recoil calibration events reported in Fig.~4 of Ref.~\cite{Akerib:2013tjd}, with an expected number of background events in that region of $0.64\pm0.16$ events. We include this information in our analysis adding a term of the form (\ref{Like_eff}) to the total likelihood, with $k=1$, $\hat{\mu}_B=0.64$ and $\sigma_B=0.16$. We calculate $\mu_S(m_\chi,\mathbf{c},\bfeta)$ integrating Eq.~(\ref{xenon100}) between 2 and 30 PE, and assuming $\sigma_{\rm PMT}=0.37$, an exposure of $250\times85.3$ kg-days, and the experimental efficiency reported in Fig.~9 of Ref.~\cite{Akerib:2013tjd}, multiplied by an additional factor 1/2, corresponding to the 50\% nuclear recoil acceptance quoted by the LUX collaboration.

\subsection{COUPP}
\label{COUPP}
The Chicagoland Observatory for Underground Particle Physics (COUPP) experiment seeks to observe bubble nucleations arising from dark matter scattering in a superheated liquid. The latest results from a 4.0 kg CF$_3$I bubble chamber operating from September 2010 to August 2011 at the SNOLAB deep underground laboratory have been reported in Ref.~\cite{Behnke:2012ys}. For this experimental apparatus, the probability that an energy $E_{R}$ nucleates a bubble above a threshold energy $E_{\rm th}$ is given by~\cite{Behnke:2012ys}
\begin{equation}
\mathcal{P}_{T}(E_{R},E_{\rm th}) = 1 - \exp\left[ -\alpha_{T} \frac{E_{R}-E_{\rm th}}{E_{\rm th}}\right] \,.
\end{equation}
This probability depends on the target nucleus. The constant $\alpha_T$, with $T=$C,F,I, is determined by fitting the above expression to the observed rates of single, double, triple and quadruple bubble events in test runs performed using neutron sources. We calculate the expected number of dark matter scattering events with an energy larger than $E_{\rm th}$ in the COUPP detector as follows~\cite{Behnke:2012ys} 
\begin{equation}
\mu_{S}(m_\chi,\mathbf{c},\bfeta) =\epsilon(E_{\rm th}) \sum_{T={\rm C,F,I}} \int_{E_{\rm th}}^{\infty}dE_{R} \, \mathcal{P}_T(E_{R},E_{\rm th}) \frac{d\mathcal{R}_{T}}{dE_{R}}
\label{coupp}
\end{equation}
where $\epsilon(E_{\rm th})$ is the threshold dependent experimental exposure multiplied by the bubble detection efficiency. The values relevant for the present analysis are $\epsilon(7.8~{\rm keV})=55.8$ kg-days, $\epsilon(11~{\rm keV})=70$ kg-days and $\epsilon(15.5~{\rm keV})=311.7$ kg-days~\cite{Behnke:2012ys}. COUPP operated in three different experimental configurations, corresponding to bubble nucleation threshold energies of 7.8 keV$_{\rm nr}$, 11 keV$_{\rm nr}$ and 15.5 keV$_{\rm nr}$ respectively. With these threshold energies, COUPP observed $k=2$, $k=3$ and $k=8$ events respectively for each experimental configuration. The corresponding estimated number of background events associated with $\alpha$-decays in the materials surrounding the CF$_3$I volume is $\mu_B=0.8$, $\mu_B=0.7$ and $\mu_B=3$ events, respectively. For each threshold energy, we add a term of type (\ref{Like_eff}) to the total likelihood if $\mu_{S}(m_\chi,\mathbf{c},\bfeta)+\mu_B>k$, and a large negative constant otherwise, which corresponds to considering the value of $k$ as an upper bound only. In all cases we set $\sigma_B=0$. 
Regarding the parameter $\alpha_i$, following \cite{DelNobile:2013sia}, we assume $\alpha_{\rm I}\rightarrow +\infty$ (i.e., perfect efficiency for bubble nucleation), $\alpha_{\rm C}=0$ (i.e., no bubble nucleation) and finally, $\alpha_{\rm F}\equiv a_{\rm COUPP}$, treating the latter as a nuisance parameter with a Gaussian prior, as shown in Tab.~\ref{prior}. We do not consider dark matter scattering on Carbon, since for this nucleus nuclear form factors are not available for all the nuclear responses emerging in the effective theory of Ref.~\cite{Anand:2013yka}.

\subsection{PICASSO}
\label{PICASSO}
The PICASSO experiment searches for dark matter using superheated liquid droplets made of C$_4$F$_{10}$~\cite{Archambault:2012pm}. In the last run, PICASSO operated in eight different experimental configurations, corresponding to the following bubble nucleation threshold energies (in keV$_{\rm nr}$): 1.7, 2.9, 4.1, 5.8, 6.9, 16.3, 38.8 and 54.8. For each energy, the collaboration reported the observed rate $\hat{\mathcal{R}}_i$ (for $i=1,\dots,8$) of dark matter scattering events above threshold including the associated experimental errors $\sigma_i$ (see Fig.~5 of Ref.~\cite{Archambault:2012pm}). We calculate the expected scattering rate $\mathcal{R}_i$ in the energy range $[E_{\rm th},+\infty]$ using Eq.~(\ref{coupp}) but setting $\epsilon=1$ and assuming $\alpha_{\rm F}=a_{\rm PICASSO}$, where $a_{\rm PICASSO}$ is the nuisance parameter described in Tab.~\ref{prior}. Similarly to the COUPP experiment, we focus on dark matter scattering off fluorine only, since carbon form factors are not available for all the nuclear responses proposed in Ref.~\cite{Anand:2013yka}. The PICASSO contribution to the total likelihood is then obtained assuming a Gaussian likelihood function for the eight PICASSO data points, namely   
\begin{equation}
-\ln \mathcal{L}_{\rm PICASSO}  = \sum_{i=1}^{8} \frac{1}{2\sigma_{i}^{2}} \left[  \mathcal{R}_i - \hat{\mathcal{R}}_i \right]^2 \,.
\end{equation}

\subsection{SIMPLE}
\label{SIMPLE}
Similarly to COUPP and PICASSO, the Superheated Instrument for Massive ParticLe Experiments (SIMPLE) uses superheated liquid detectors made of C$_2$ClF$_5$ to search for bubble nucleations produced by dark matter scattering in the detector volume~\cite{Felizardo:2011uw}. We focus here on the Stage 2 data, collected with an effective experimental exposure of 6.71 kg-days, after cutting data obtained at pressures greater than 2.20 bar. In this run the experiment operated with a bubble nucleation threshold energy of 8 keV$_{\rm nr}$. Analyzing these data, the SIMPLE collaboration observed 1 candidate dark matter event. This result is consistent with an expected number of background events equal to $2.2\pm 0.3$. To estimate the expected number of dark matter scattering events in the SIMPLE detector, we use an equation analogous to Eq.~(\ref{coupp}), with the parameter $\alpha_{\rm F}\equiv a_{\rm SIMPLE}$ treated as a nuisance parameter with a Gaussian prior, as shown in Tab.~\ref{prior}. To use the density matrix approach of Ref.~\cite{Anand:2013yka} in the calculation of the nuclear form factors, we restrict our analysis to dark matter scattering off fluorine (i.e., we set $\alpha_{\rm Cl}=\alpha_{\rm C}=0$). The SIMPLE contribution to the total likelihood is then of type (\ref{Like_eff}) with $k=1$, $\hat{\mu}_B=2.2$ and $\sigma_B=0.3$.

\subsection{DAMA}
DAMA uses highly-radiopure thallium-doped NaI scintillators to detect dark matter through the observation of an annual modulation in the measured nuclear recoil energy spectrum of the target sodium and iodine nuclei~\cite{Bernabei:2010mq}. The modulation signal is expected as a consequence of the Earth's motion through the Milky Way dark matter halo, which sinusoidally modulates the flux of dark matter particles impinging on the DAMA detector, with a period of one year~\cite{2012arXiv1209.3339F}. In the conventional dark halo model, the modulation is expected to be at a maximum on June 2 and at a minimum on December 2. Combining the data collected over 7 annual cycles by the DAMA/NaI configuration of the experiment with the data of the 6 annual cycles recorded by its upgrade DAMA/LIBRA,  the total exposure of the DAMA experiment reaches 1.17 ton$\times$year. As for any direct detection experiment relying on scintillators, only a certain fraction of the true nuclear recoil energy deposited by dark matter particles in the DAMA NaI crystals is directly accessible to the photomultipliers installed in the experimental apparatus. The measurable scintillation energy $E_{\mathcal{O}}$ (in keV electron equivalent units, i.e., keV$_{\rm ee}$) is related to the true nuclear recoil energy $E_{R}$ (in keV nuclear recoil units, i.e., keV$_{\rm nr}$, or simply keV) by a quenching factor, denoted by $q_{\rm Na}$ and $q_{\rm I}$ for sodium and iodine nuclei, respectively. The value of the quenching factor is uncertain, and different choices have been explored in the literature. In our fits we set $q_{\rm I}=0.09$, and treat $q_{\rm Na}$ as a nuisance parameter varying with a Gaussian prior in the range shown in Tab.~\ref{prior}. In the latter case $E_{\mathcal{O}}=q_{\rm Na} E_{R}$. To evaluate Eq.~(\ref{rate_obs}) we also need the DAMA energy resolution, which is
\begin{equation}
\sigma(E_{\mathcal{O}}) = 0.448 \sqrt{E_{\mathcal{O}}/{\rm keV}_{\rm ee}} +9.1\times 10^{-3} E_{\mathcal{O}}/{\rm keV}_{\rm ee}\,.
\end{equation}
In this paper we use the annual modulation data reported in Fig.~6 of Ref.~\cite{Bernabei:2010mq}, including the first 12 energy bins only, since for energies larger than 8 keV$_{\rm ee}$ DAMA has not observed any statistically significant modulation effect. For these data we assume the likelihood function
\begin{equation}
-\ln\mathcal{L}_{\rm DAMA} = \sum_{i=1}^{12}\frac{1}{2\sigma_{i}^{2}}\left[S_{\rm m}(\hat{E}^{i}_{\mathcal{O}})-\hat{S}_{\rm m}(\hat{E}^{i}_{\mathcal{O}})\right]^2 ,
\end{equation}
where the expected annual modulation amplitude as a function of the energy bin lower bound $\hat{E}^{i}_{\mathcal{O}}$ is calculated as follows~\cite{2012arXiv1209.3339F}
\begin{equation}
S_{\rm m}(\hat{E}^i_{\mathcal{O}}) = \frac{1}{2\Delta \hat{E}_{\mathcal{O}}} \int_{\hat{E}^i_{\mathcal{O}}}^{\hat{E}^i_{\mathcal{O}}+\Delta\hat{E}_{\mathcal{O}}}
d\hat{E}_{\mathcal{O}} 
\left(\frac{d\mathcal{R}}{d\hat{E}_{\mathcal{O}}}\bigg|_{\rm June} - \frac{d\mathcal{R}}{d\hat{E}_{\mathcal{O}}}\bigg|_{\rm December}\right) \,.
\end{equation}
Here $\sigma_i$ are the errors associated with the 12 datapoints $\hat{S}_{\rm m}(\hat{E}^{i}_{\mathcal{O}})$ and $\Delta \hat{E}_{\mathcal{O}}=0.5$ keV$_{\rm ee}$ is the width of the energy bins.

\subsection{CoGeNT}
The CoGeNT experiment searches for a dark matter signal employing p-type point contact germanium detectors. After 1136 live days of data taking, the collaboration has recently published a new measurement of the observed nuclear recoil energy spectrum and of its time dependence~\cite{Aalseth:2014jpa}. The data show an exponential-like irreducible background of events that cannot be associated with cosmogenically activated nuclei decaying via L-shell or K-shell electron capture. In addition, the data collected also provide moderate evidence in favor of a modulation signal, with a phase subject to large uncertainties implying a peak date of $t_{\rm max}=106\pm24$ days. Using the same data, Ref.~\cite{Aalseth:2014eft} has found that the ratio between the modulation amplitude and the total unmodulated signal (i.e., the fractional amplitude) is equal to $\hat{\mathcal{A}}=(12.4\pm5)$\%, with a dark matter signal estimated to be $35\%$ of the total unmodulated signal (see also Ref.~\cite{Davis:2014bla} for an independent analysis of the CoGeNT 2014 data).
For the CoGeNT experimental apparatus, the relation between the observable energy and the true nuclear recoil energy is $E_{\mathcal{O}}=0.199\times E_{R}^{1.12}$.  In the fits we employ this relation and the recoil energy spectrum extracted from Fig.~10 in Ref.~\cite{Aalseth:2014jpa} as follows. We find the observed energy spectrum $d\hat{\mathcal{R}}/d\hat{E}_{\mathcal{O}}$ at the energy $\hat{E}^{i}_{\mathcal{O}}$ by subtracting the best fit background estimate in Fig.~10 to the observed datapoints (labeled here by an index $i$). We denote by $\sigma_i$ the error associated to this spectrum. When fitting these data, we have set $\mathcal{E}=1$ and $\sigma\rightarrow 0$ in Eq.~(\ref{rate_obs}). For this data sample we have assumed a multivariate Gaussian contribution to the total likelihood, namely
\begin{equation}
-\ln \mathcal{L}_{\rm CoGeNT}  = \sum_{i=1}\frac{1}{2\sigma_{i}^{2}} \left[\frac{d\mathcal{R}}{d\hat{E}_{\mathcal{O}}}(\hat{E}^{i}_{\mathcal{O}})-\frac{d\hat{\mathcal{R}}}{d\hat{E}_{\mathcal{O}}}(\hat{E}^{i}_{\mathcal{O}})\right]^2 + \frac{1}{2\sigma_{\mathcal{A}}^2} (\mathcal{A}_{\rm theory}-\hat{\mathcal{A}})^2,
\end{equation}
Here $\sigma_{\mathcal{A}}=5$\%, and the second term in this expression takes into account the information on the CoGeNT annual modulation. When calculating the expected fractional amplitude $\mathcal{A}_{\rm theory}$, we have treated the peak date $t_{\rm max}$ as a nuisance parameter with a Gaussian prior as shown in Tab.~\ref{prior}.

\section{Limits on the dark matter-nucleon interaction strength}
\label{result1}
We now compare the predictions of the dark matter theory introduced in Sec.~\ref{theory} to the data of the previous section, using the statistical tools summarized in Sec.~\ref{statistics}. We assume for definiteness that the dark matter particle has spin $j_\chi=1/2$. In this section, we  focus on experiments compatible with a null result. In the next section we study experiments with a dark matter signal, using the same theoretical and statistical frameworks. 

\subsection{Limits from single experiments}
\label{sec:LUX}
\begin{figure}[t]
\centering
\includegraphics[width=\textwidth]{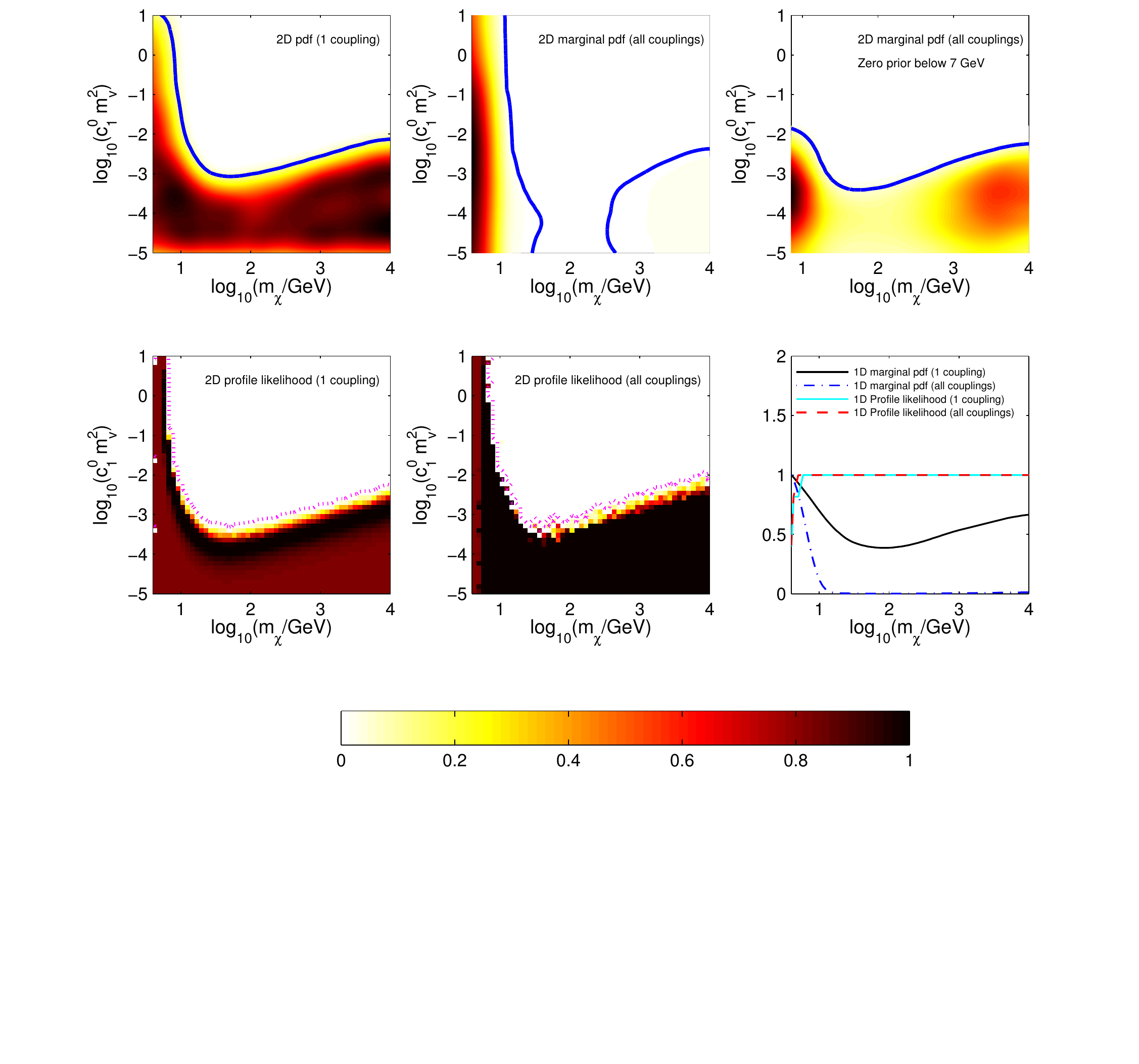}
\vspace{-4 cm}
\caption{Analysis of the LUX data. {\it Top-left panel}: 2D posterior PDF and associated 99\% credible region (CR) in the $m_{\chi}$--$c_1^{0}$ plane, obtained by fitting the LUX data varying $m_{\chi}$ and $c_1^0$ only (in this analysis the remaining couplings have been set to zero). Below the 99\% CR contour, the marginal posterior PDF increases and it reaches a plateau of approximately constant probability. {\it Top-central panel}: 2D marginal posterior PDF and 99\% credible regions in the $m_{\chi}$--$c_1^{0}$ plane, obtained by fitting the LUX data varying $m_{\chi}$ and all the effective couplings simultaneously. Because of volume effects (see text around Eqs.~(\ref{eq:prof_likelihood}) and (\ref{approxmarg})) the marginal posterior PDF is now peaked at low masses and the 99\% CR splits into two islands, one at low masses and the other at high masses. {\it Top-right panel}: As for the top-central panel, but with zero prior below 7 GeV for $m_{\chi}$. Also in this case, the 2D marginal posterior PDF peaks at low masses. {\it Bottom-left panel}: 2D profile likelihood in the $m_{\chi}$--$c_1^{0}$ plane, extracted from the LUX data varying $m_{\chi}$ and $c_1^{0}$ only. Below the 95\% Confidence Level (CL) contour, the profile likelihood increases, it reaches a region of maxima, and then it decreases assuming a constant value corresponding to $\mu_S(m_\chi,\mathbf{c},\bfeta)=0$.  {\it Bottom-central panel}: 2D profile likelihood in the $m_{\chi}$--$c_1^{0}$ plane, obtained from an analysis of the LUX data in which we vary $m_{\chi}$ and all the effective couplings simultaneously. Importantly, the 2D profile likelihood surface does not split into disconnected regions, since it is unaffected by volume effects. In addition, when varying more than one parameter, configurations with relatively low likelihood corresponding to $\mu_S(\mathbf{c},\eta)=0$ occur at low masses only (lighter band near the left margin of the bottom-central panel). {\it Bottom-right panel}: 1D posterior PDFs obtained marginalizing over the coupling constants, and 1D profile likelihoods associated with the four cases discussed in the other panels of this figure.}
\label{c1_LUX}
\end{figure}

We start with a detailed analysis of the LUX data, to illustrate the many physical effects and computational challenges which can be encountered when studying a parameter space of large dimensionality, like in this work. The top-left panel of Fig.~\ref{c1_LUX} shows the results of a fit where we have considered as free parameters $c_1^0$ and $m_{\chi}$ only, setting to zero all the remaining couplings. This corresponds to the standard case in which the LUX data are interpreted in terms of spin-independent interactions, with one important difference, however, namely the fact that instead of presenting Confidence Levels (CL) in the $m_{\chi}$--$\sigma_{\rm p}^{\rm SI}$ plane, where the latter is the $\chi$-proton cross section, we present the 2D posterior PDF and its associated 99\% Credible Region (CR) in the related $m_{\chi}$--$c_1^0$ plane. The connection is, from Eqs.~(\ref{dsigmadER})-(\ref{Ptot}) keeping the isoscalar part only, 
\begin{align}
\sigma_{N}^{\rm SI} = \frac{\mu_{N}^2 |c_1^0|^2}{4\pi},
\label{eq:sigmaSIp1}
\end{align}
where $\mu_{N}=m_\chi m_{N}/(m_\chi+m_{N})$ is the reduced $\chi$-nucleon mass.  For reference, the familiar spin-dependent cross section $\sigma_{N}^{\rm SD}$ is related to $c_4^0$ by
\begin{align}
\sigma_{N}^{\rm SD} = \frac{\mu_N^2 j_\chi (j_\chi+1) |c_4^0|^2}{16\pi} .
\end{align}
In the top-left panel of Fig.~\ref{c1_LUX} we observe a smooth 99\% CR contour and a posterior PDF that grows below this contour to reach a plateau of approximately constant posterior probability. The calculation to produce this plot required $\mathcal{O}(10^4)$ likelihood-function evaluations, even when additional nuisance parameters are included in the fit to model various sources of uncertainty (as done for instance in the case of XENON100). 

In a second analysis, we fit the LUX data varying the ten couplings $c_i^0$, with $i=1,3,\dots,11$, and setting to zero the analogous isovector couplings, assuming isospin-conserving interactions. The top-central panel of Fig.~\ref{c1_LUX} shows  the 2D marginal posterior PDF and the associated 99\% CR in the $m_\chi$--$c_1^0$ plane. Important differences are observed between this PDF (marginalized over 9 parameters) and the previous one in the top-left panel (with the 9 parameters set to zero). The marginalized PDF is peaked at low masses, with a 99\% CR contour consisting of two disconnected ``islands,'' one, more pronounced, at small masses, and the other at large masses. Qualitatively, the appearance of the ``islands'' can be analytically understood as a volume effect emerging during the marginalization process (see discussion around Eq.~(\ref{eq:prof_likelihood})). In fact, if correlations between different $c^0_i$ can be neglected, the 2D marginal posterior PDF $\mathcal{P}_{\rm marg}(c_i^0,m_{\chi})$ factorizes as follows
\begin{eqnarray}
\mathcal{P}_{\rm marg}(c_j^0,m_{\chi}) &\propto& \left[ \prod_{i\neq j} \int dc_i^0  \mathcal{P}^{(i)}(c^0_i | m_{\chi}) \right] \times \mathcal{P}^{(j)}(c^0_j,m_{\chi}) \nonumber\\
&\equiv&\Gamma(m_{\chi})   \mathcal{P}^{(j)}(c^0_j,m_{\chi})  .
\label{approxmarg}
\end{eqnarray}
In the first line, $\mathcal{P}^{(i)}(c^0_i | m_{\chi})  = \mathcal{P}^{(i)}(c^0_i , m_{\chi})/ \mathcal{P}^{(i)}(m_\chi)$ is the conditional probability of the $c^0_i$ given $m_\chi$. The posterior PDF $\mathcal{P}^{(j)}(c^0_j,m_{\chi})$ is the posterior PDF for $c^0_i=0$ when $i\ne j$. In our fit of the LUX data, $\mathcal{P}^{(1)}(c^0_1,m_{\chi})$ is shown in the top-left panel of Fig.~\ref{c1_LUX}. The function $\Gamma(m_\chi)$ introduced in Eq.~(\ref{approxmarg}) is very sensitive to the prior at masses $m_\chi$ where direct detection experiments are not sensitive. For our flat prior, $\Gamma(m_\chi)$ follows the  $m_\chi$ dependence of a typical direct detection exclusion limit, in the following sense: (1) it is peaked at low masses, since in this region the range of $c^0_i$ for which $\mathcal{P}^{(i)}(c^0_i | m_{\chi})\neq 0$ is infinite; (2) it has a minimum at $m_\chi\sim$50 GeV, where exclusion limits are usually stronger; (3) it is slightly less suppressed at larger $m_{\chi}$, where exclusion limits tend to be less restrictive. This behavior of $\Gamma({m_\chi})$ produces the pattern seen in the top-central panel of Fig.~\ref{c1_LUX}, where the 2D marginal PDF $\mathcal{P}_{\rm marg}(c_j^0,m_{\chi})$ is highly peaked at low masses, has a minimum at $m_\chi \sim$50 GeV, and slowly increases at larger masses. A change in the prior range of $m_{\chi}$ leads to a function $\Gamma(m_\chi)$ with a qualitatively similar behavior, as shown in the top-right panel of the same figure. The volume effects are however uncomfortable.

As already mentioned, a statistical indicator insensitive to the volume effects is the profile likelihood. The bottom-left panel of Fig.~\ref{c1_LUX} shows the 2D profile likelihood, together with the associated 95\% CL contour, obtained from a fit of the LUX data where we vary $c_1^0$ and $m_\chi$ only. Below the 95\% CL contour, the profile likelihood increases, it reaches a region of maxima corresponding to an expected count rate $\mu_S(m_\chi,\mathbf{c},\bfeta)+\mu_{B}\sim 1$, and finally, it decreases towards a plateau associated with $\mu_S(m_\chi,\mathbf{c},\bfeta)=0$, and therofore with $\mu_S(m_\chi,\mathbf{c},\bfeta) + \mu_B = 0.64$. Importantly, when varying a single coupling, CL contours and CR contours agree well, in regions where the latter make sense. The bottom-central panel shows the 2D profile likelihood extracted from a fit of the LUX data performed  varying all the couplings and the dark matter mass simultaneously. Contrary to the case in which a single coupling is varied, the region of maxima extends everywhere below the 95\% CL contour, except at low masses, where the expected dark matter signal is below the experimental threshold and $\mu_S(m_\chi,\mathbf{c},\bfeta)=0$ independently of $\mathbf{c}$. The presence of an infinite plateau of maxima can be explained as follows: even when $c_1^0$ is very small, the value of  $\mu_S(m_\chi,\mathbf{c},\bfeta)$, which would be tiny if all the other couplings were zero, can be sufficiently large to satisfy the maximum condition $\mu_S(m_\chi,\mathbf{c},\bfeta)+\mu_{B}\sim 1$, because of contributions associated with the other couplings. Only at sufficiently low values of $m_\chi$, $\mu_S(m_\chi,\mathbf{c},\bfeta)=0$ independently of $\mathbf{c}$ because of the already mentioned threshold effects. 
\label{singlefit}
\begin{figure}[t]
\begin{center}
\includegraphics[width=\textwidth]{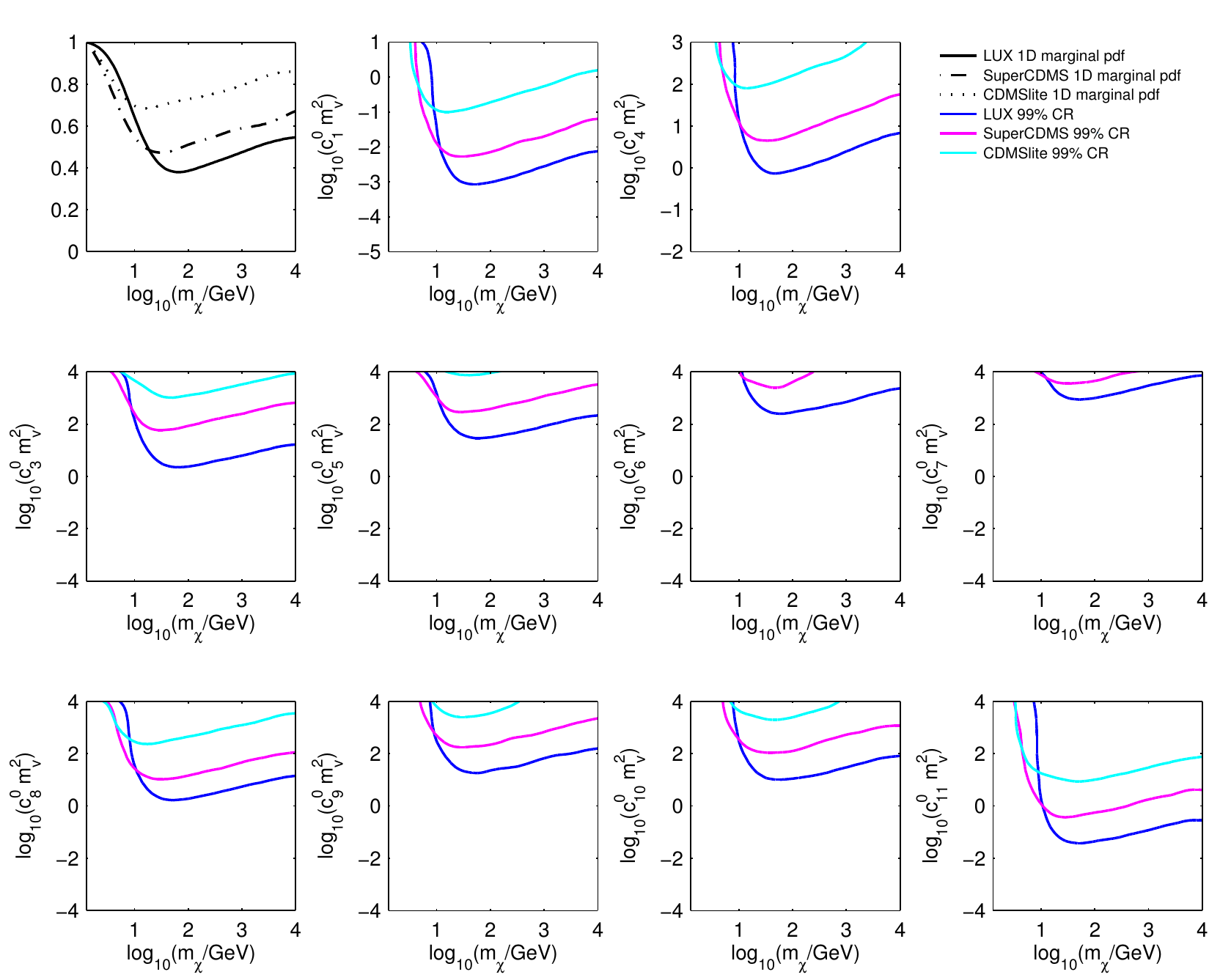}
\end{center}
\caption{Bayesian analysis of the LUX, SuperCDMS and CDMSlite data.  {\it Top-left panel}: 1D marginal posterior PDF of the dark matter mass extracted from the three datasets. These lines are obtained by sampling the posterior PDF varying $m_{\chi}$ and $c_1^0$ only, and then marginalizing over $c_1^0$. In all cases this 1D PDF is at its maximum for small $m_{\chi}$. This reflects the fact that for small $m_{\chi}$ there is a larger fraction of the parameter space that is allowed by the data (similar results are obtained when varying one of the other couplings $c^0_i$, $i\ne1$). {\it Other panels}: 99\% CR contours extracted from the three datasets by varying $m_{\chi}$ and one single coupling only (different panels correspond to distinct couplings).}
\label{SL1}
\end{figure}
\begin{figure}[t]
\begin{center}
\includegraphics[width=\textwidth]{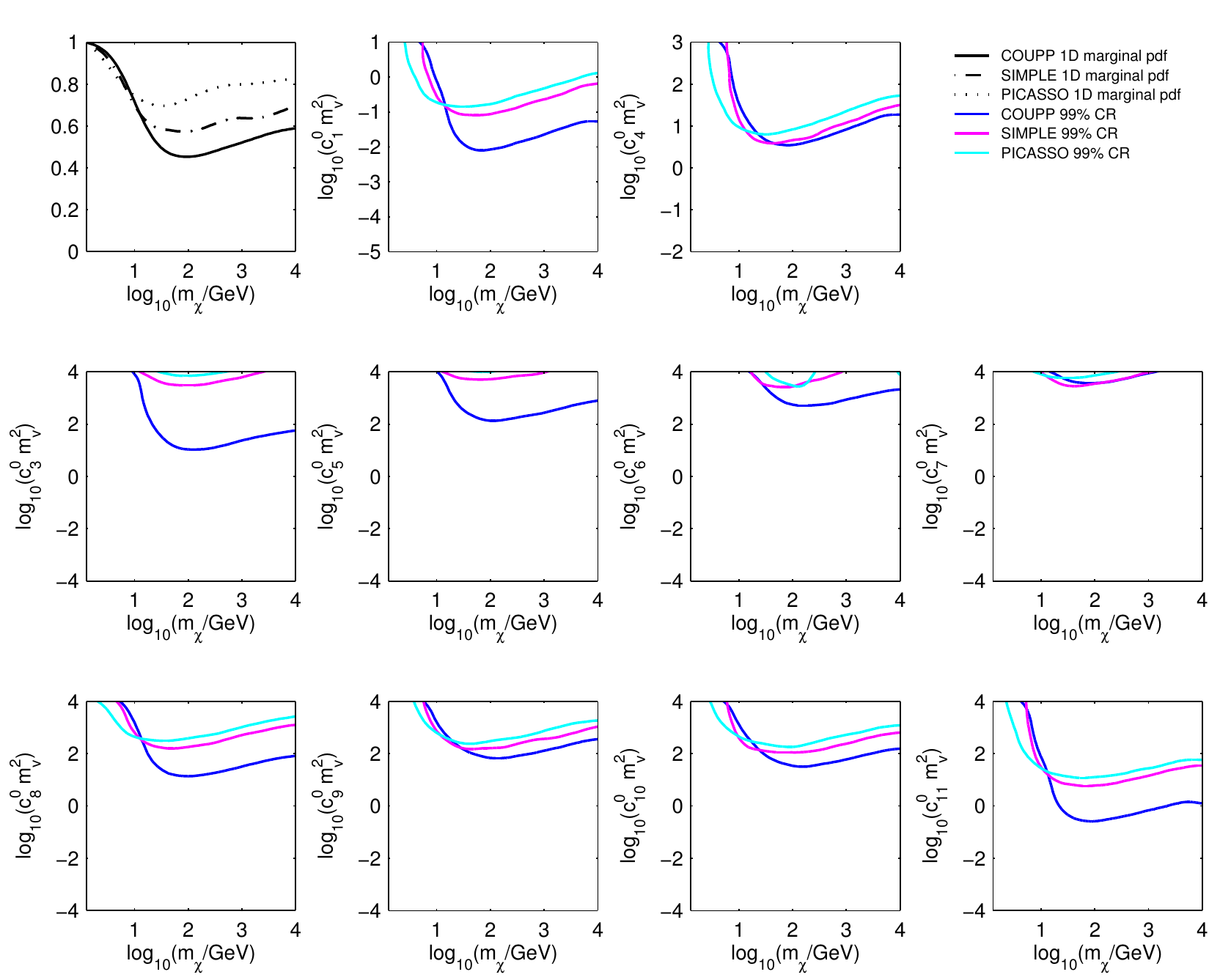}
\end{center}
\caption{Same as Fig.~\ref{SL1}, but for the PICASSO, COUPP, and SIMPLE data.}
\label{SL2}
\end{figure}
\begin{figure}[t]
\begin{center}
\includegraphics[width=\textwidth]{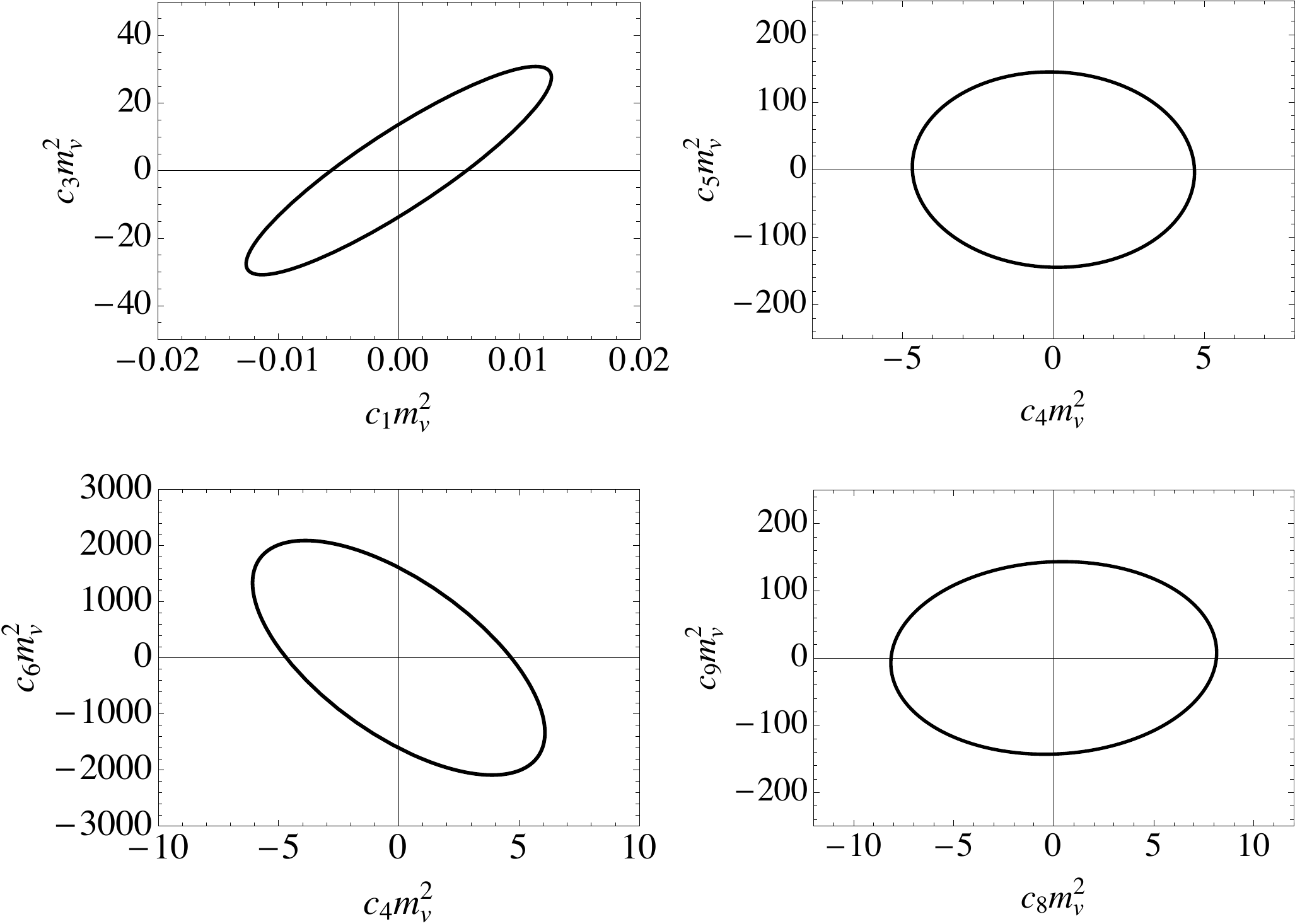}
\end{center}
\caption{95\% CL profile-likelihood upper limits on the coupling constants $c_i^0$ ($i=1,3,\ldots,11$) that can in principle exhibit correlations, for the LUX experiment and a dark matter particle mass $m_\chi = 10 $ TeV. There is negligible correlation between $c_4^0$ and $c_5^0$ and between $c_8^0$ and $c_9^0$, positive correlation  between $c_1^0$ and $c_3^0$, and negative correlation  between $c_4^0$ and $c_6^0$.}
\label{cicj}
\end{figure}
\begin{figure}[t]
\begin{center}
\includegraphics[width=1.05\textwidth]{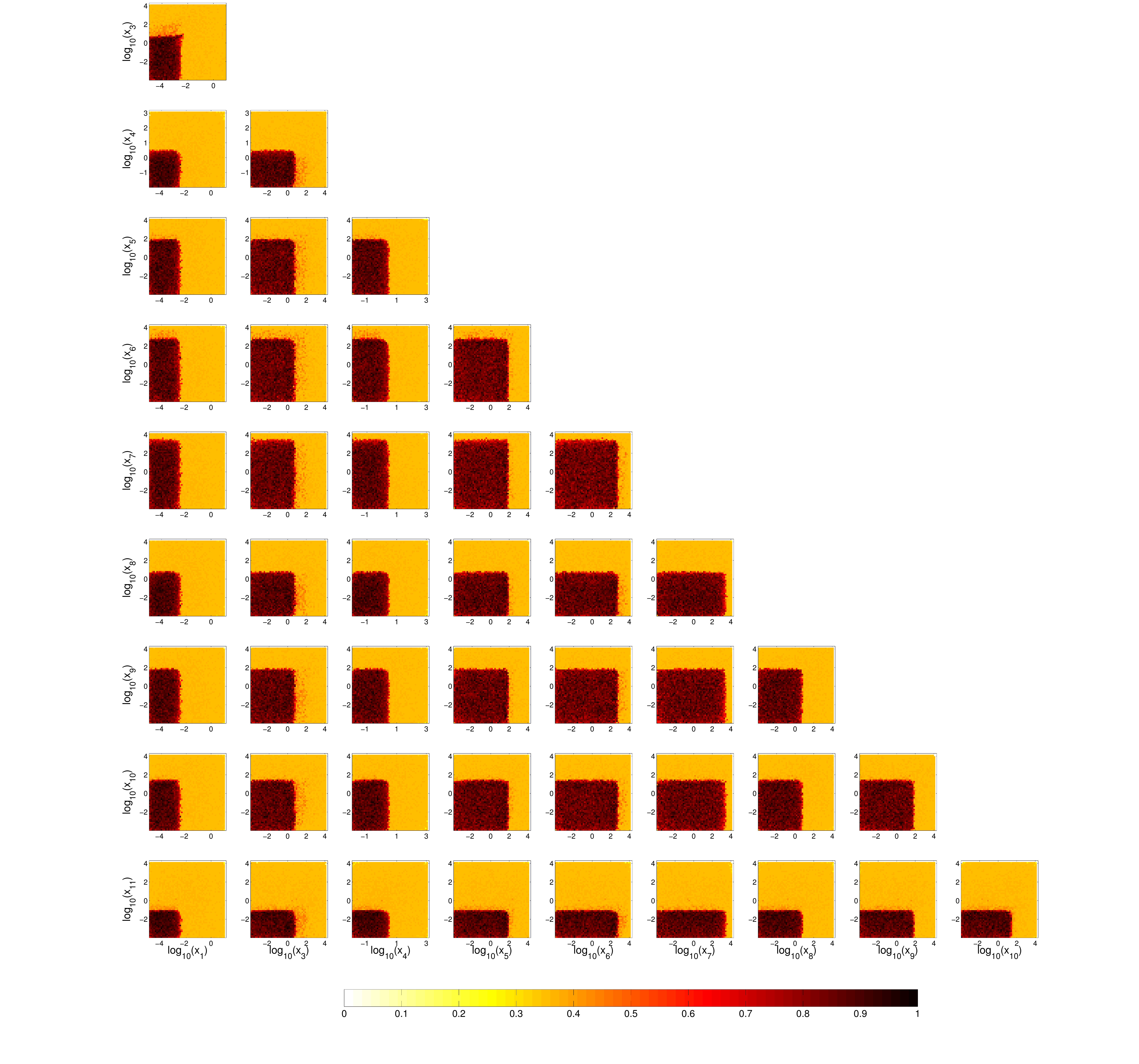}
\end{center}
\caption{2D profile likelihood in the 45 planes spanned by all the independent pairs of effective couplings considered in this work. For illustrative purposes we have introduced in this figure the new variables $x_i\equiv c_i^0 m_v^2$, with $i=1,3,\dots,11$. These 2D profile likelihoods have been extracted from an analysis in which all the datasets with  null results were fit simultaneously varying all the effective couplings and the dark matter mass (together with the nuisance parameters). This figure clearly shows the absence of strong correlations between the different effective couplings, except between $c_1^0$--$c_3^0$ and $c_4^0$--$c_6^0$ (see text and Figs.~\protect\ref{cicj} and \protect\ref{c1c3}).}
\label{10cs_corr}
\end{figure}
\begin{figure}[t]
\begin{center}
\includegraphics[width=0.7\textwidth]{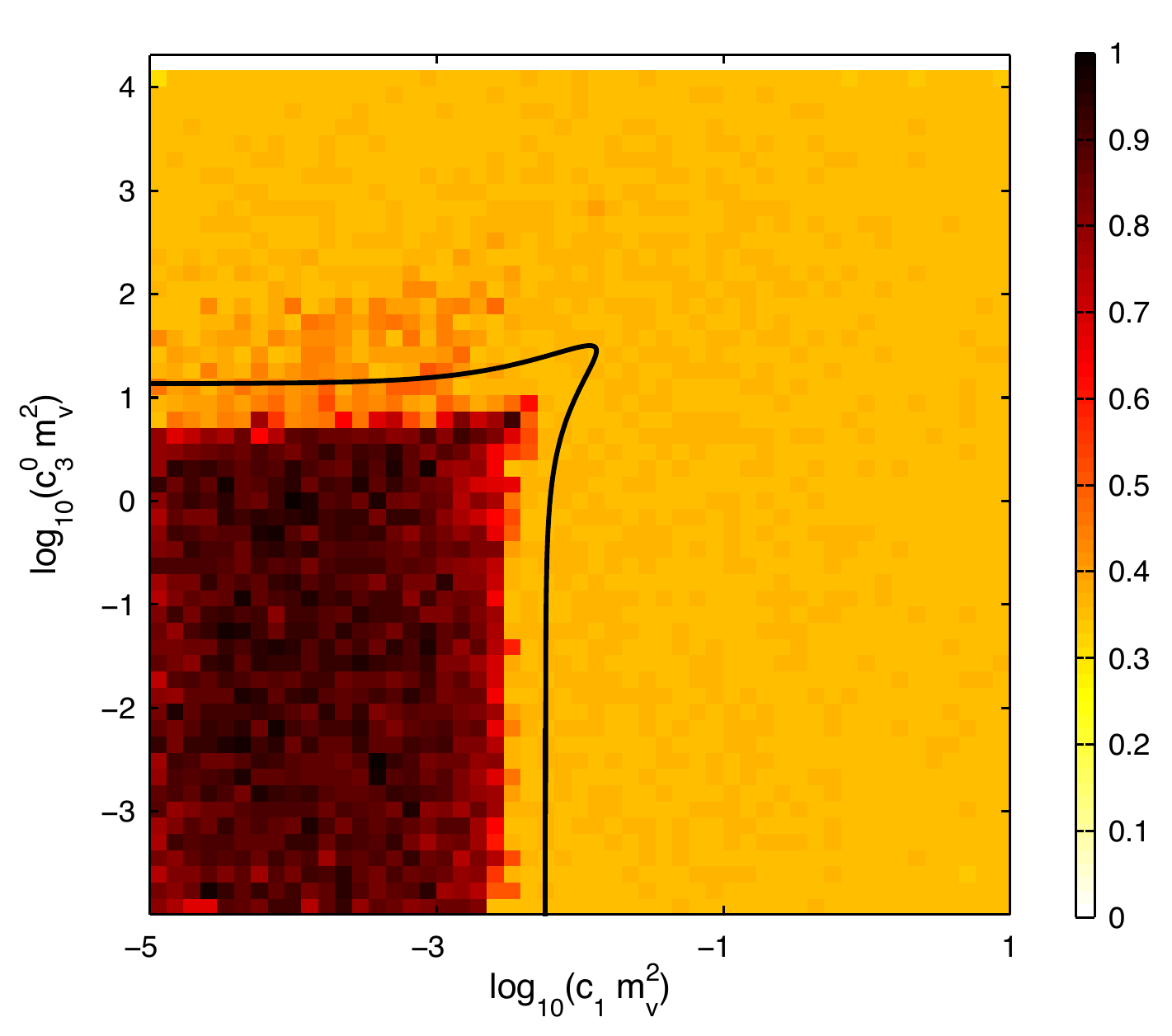}
\end{center}
\caption{{\it Color scale:} 2D profile likelihood in the $c_1^0$--$c_3^0$ plane from a global analysis of all datasets with null results (enlargement of the top-left panel in Fig.~\protect\ref{10cs_corr}). {\it Black line:} log-log graph of the ellipse in Fig.~\protect\ref{cicj}, representing the 95\% CL upper limit from LUX at $m_\chi=10$ TeV. The positive correlation between $c_1^0$ vs $c_3^0$ is embodied in the feature that protrudes at the corner of the dark region following the black line.}
\label{c1c3}
\end{figure}
\begin{figure}[t]
\begin{center}
\includegraphics[width=\textwidth]{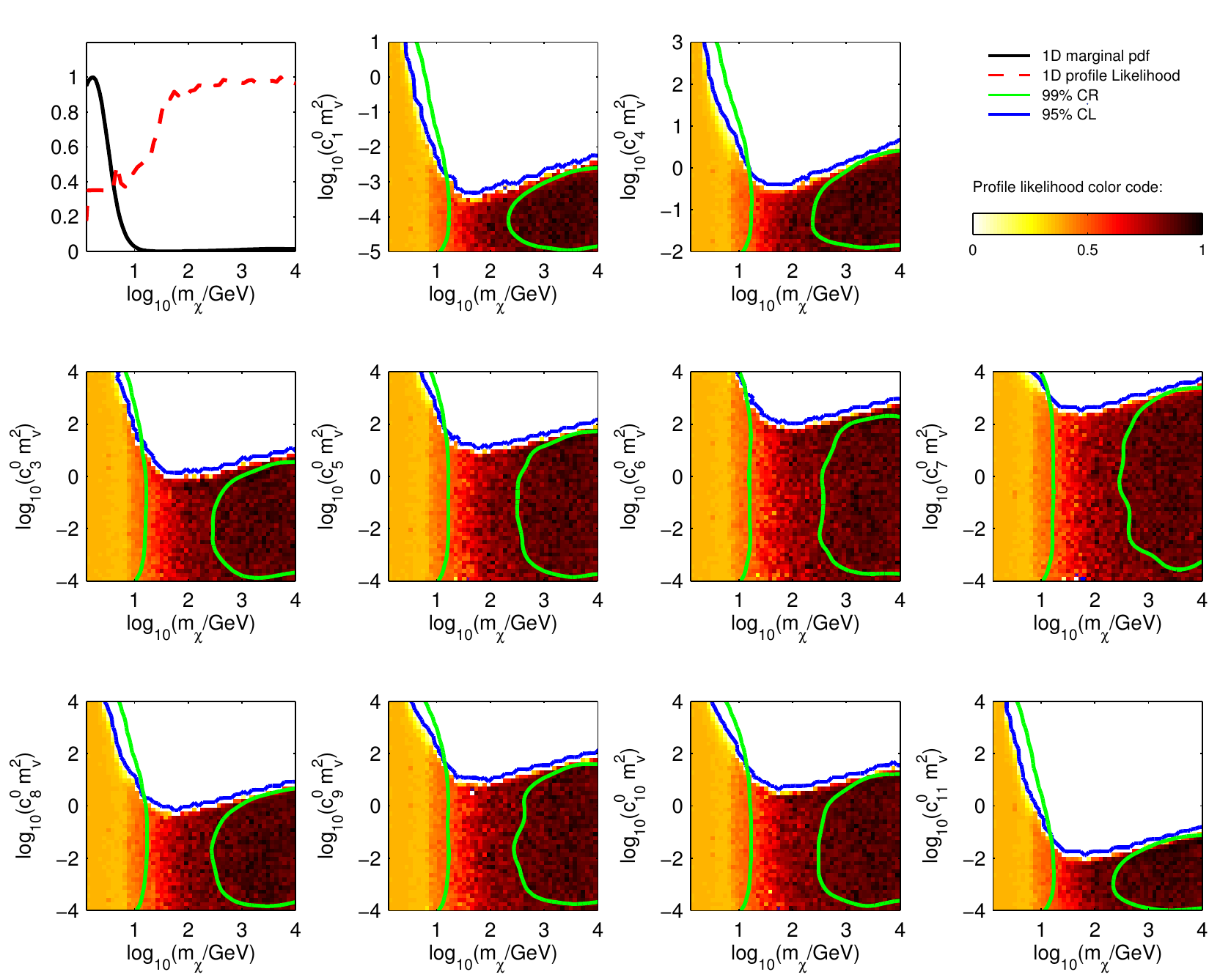}
\end{center}
\caption{This figure illustrates the main result of this work. In 10 planes spanned by $m_{\chi}$ and one of the effective couplings we show the corresponding 99\% CR contours (green), the 95\% CL contours (blue) and the associated 2D profile likelihoods. These statistical indicators have been constructed through a global fit of all the datasets considered in this work (except DAMA and CoGeNT) in which we have simultaneously varied the dark matter mass, all the effective couplings and the nuisance parameters introduced in the previous sections. From the 95\% CL contours in this figure one can extract the maximum strength compatible with current direct detection data of the different types of dark matter-nucleon interaction as a function of the dark matter mass. The top-left panel shows the 1D marginal PDF and the 1D profile likelihood of the dark matter mass resulting from this global analysis. The 1D PDF is suppressed at large $m_{\chi}$ because of volume effects, whereas the 1D profile likelihood is lower at small $m_{\chi}$ because of threshold effects (see text).}
\label{10cs}
\end{figure}

The bottom-right panel of Fig.~\ref{c1_LUX} illustrates the two 1D posterior PDFs obtained marginalizing over the coupling constants, and the two 1D profile likelihoods associated with the four CR/CL contours described in the previous paragraphs. This figure shows the flatness of the profile likelihoods (modulo threshold effects) and the location of the peaks of the 1D marginal posterior PDFs.

The conclusions, illustrated here in detail in the important case of the LUX experiment, also apply - of course with obvious quantitative differences - to the other experiments considered in this paper, and to coupling constants different from $c_1^0$. For instance, Figs.~\ref{SL1} and \ref{SL2} show the 99\% CR contours in the 10 planes $m_\chi$ vs $c^0_i$ obtained from the LUX, SuperCDMS, CDMSlite, COUPP, SIMPLE and PICASSO data.\footnote{To keep the figures simple, we do not include XENON10, XENON100, CDMS-Ge, and CDMS-LT.} To extract these contours we analyze each dataset independently and sample the posterior PDF by varying $m_\chi$  and one of $c^0_i$ at a time, in addition to the relevant nuisance parameters. We have verified that the limits obtained in this way in the planes $m_\chi$ vs $c_1^0$ and $m_\chi$ vs $c_4^0$, match very well standard results usually presented in the planes $m_\chi$ vs $\sigma_{n}^{\rm SI}$ and $m_\chi$ vs $\sigma_{n}^{\rm SD}$. In addition, from Figs.~\ref{SL1} and \ref{SL2} one can also extract the maximum strength allowed by these experiments for couplings different from the familiar $\mathcal{O}_1$ and $\mathcal{O}_4$ interactions. Notably, experiments such as COUPP and LUX are able to set important constraints on all the interactions types considered in this paper. The interaction $\mathcal{O}_1$ is the most severely constrained (at the level of $c_i^0 m_v^2 \lesssim 10^{-3}$ for $m_\chi \sim 50$ GeV), followed by the interactions $\mathcal{O}_{11}$ (at the level of $c_i^0 m_v^2 \lesssim 5\times 10^{-2}$) and $\mathcal{O}_3$, $\mathcal{O}_4$, $\mathcal{O}_8$ (at the level of $c_i^0 m_v^2 \lesssim 1$). Within the two groups of experiments in Figs.~\ref{SL1} and \ref{SL2} the sensitivity ranking observed in the case of the couplings  $c_1^0$ and $c_4^0$ is also found for the other types of interactions (for instance $c_8^0$ and $c_{11}^0$): CDMSlite and PICASSO set the most stringent limits at very low masses, SuperCDMS and SIMPLE tend to be the leading experiments in a small window at low masses, and LUX and COUPP dominate above $m_{\chi}\sim 10$ GeV or so.

 \subsection{Global limits}
So far we have derived constraints on the couplings $c_i^0$ analyzing different direct detection experiments separately. We now move to the more complex problem of combining the complementary information contained in different direct detection experiments. More specifically, we focus here on experiments which led to a null result. The DAMA and CoGeNT experiments, which claim a signal, are discussed in the next section. 

Before describing our numerical results, we present some semi-analytic considerations on the correlation between the coupling constant $c_i^0$. The expected number of dark matter events $\mu_S(m_\chi,\mathbf{c},\bfeta)$ is a quadratic function of the constants $c_i^0$, as follows from Eqs.~(\ref{Ptot}), (\ref{rate_theory}), (\ref{rate_obs}), (\ref{eq:muS}), and (\ref{eq:Rfunctions}). As a consequence, the likelihood function in Eq.~(\ref{Like}) at fixed $m_\chi$ and $\bfeta$ is constant on ellipsoids in the coupling constants $c_i^0$. Correlations between $c_i^0$ and $c_j^0$,  with $i\ne j$,  arise from the cross term $c_i^0 c_j^0$ in $\mu_S(m_\chi,\mathbf{c},\bfeta)$. Inspection of the dark matter response functions $R^{\tau\tau'}$ in Eq.~(\ref{eq:Rfunctions}) shows that the only cross terms, and thus the only possible correlations, among the 10 coupling constants we consider are between $c_1^0$ and $c^3_0$, $c_4^0$ and $c_5^0$, $c_4^0$ and $c_6^0$, and $c_8^0$ and $c_9^0$. If we consider the correlation for one of these pairs ($c_i^0$ and $c_j^0$, say) setting the other coupling constants to zero, the contours of constant $\mu_S(m_\chi,\mathbf{c},\bfeta)$ for a given experiment at a given $m_\chi$ and $\bfeta$ are ellipses in the $c_i^0$--$c_j^0$ plane. These ellipses can be obtained without random sampling in parameter space by writing
\begin{align}
a_{ii} (c_i^0)^2 + 2 a_{ij} c_i^0 c_j^0 + a_{jj} (c_j^0)^2 =\mu_{S\rm const},
\label{eq:ellipses}
\end{align}
where $\mu_{S\rm const}$ is the desired value of $\mu_S$ (e.g., its upper limit) and the coefficients $a_{ii}$, $a_{ij}$, and $a_{jj}$ are obtained using Eqs.~(\ref{Ptot}), (\ref{rate_theory}), (\ref{rate_obs}), (\ref{eq:muS}), and (\ref{eq:Rfunctions}). The relative size of these coefficients, and thus the shape of the ellipses, is essentially fixed by the nuclear structure functions $W$. The correlation coefficient $r_{ij}$ for the pair of variables $c_i^0$ and $c_j^0$ follows as
\begin{align}
r_{ij} = - \frac{a_{ij}}{\sqrt{a_{ii} a_{jj}}}.
\end{align}
Fig.~\ref{cicj} shows the ellipses (\ref{eq:ellipses}) for LUX at $m_\chi = 10 $ TeV, with $\mu_{S\rm const}$ corresponding to the LUX upper limit. We see that out of the four possible cases, two exhibit negligible correlations (with $r_{45}=-0.027$ and $r_{89}=0.054$), one has positive correlation ($c_1^0$ and $c_3^0$ with $r_{13}=0.90$) and one has negative correlation ($c_4^0$ and $c_6^0$ with $r_{46}=-0.64$). These correlations survive when all experiments are included in the profile likelihood analysis, as seen next.

We exploit the {\sffamily Multinest} program to explore the multidimensional parameter space of the dark matter-nucleon effective theory by simultaneously varying the 11 model parameters and the 4 additional nuisance parameters listed in Tab.~\ref{prior}. Our analysis is based on about 3 million likelihood evaluations. 

Fig.~\ref{10cs_corr} shows the 2D profile likelihoods in the planes $c_i^0$ vs  $c_j^0$ (with $i,j=1,3,\dots,11$ and $i\neq j$), obtained by profiling out all parameters but $c_i^0$ and  $c_j^0$. There are 45 independent pairs of the 10 coupling constants $c_i^0$, leading to the 45 panels in Fig.~\ref{10cs_corr}. In spite of the repetitiveness of these plots, this figure contains a very important result: as shown by the absence of preferred directions in the 2D profile likelihoods, there are no evident correlations induced by the data between most pairs of the 10 couplings $c_i^0$ (except for $c_1^0$--$c_3^0$ and $c_4^0$--$c_6^0$), as expected from the semi-analytic considerations at the beginning of this section. Using the 2D marginal posterior PDFs in place of the profile likelihoods leads to an identical conclusion. The correlation between $c_1^0$ and $c_3^0$ is evidenced by the small ``spur'' protruding from the corner of the dark region in the top-left panel in Fig.~\ref{10cs_corr}, which is enlarged in Fig.~\ref{c1c3} . To wit, the black line in Fig.~\ref{c1c3} is the graph of an ellipse in a log-log plane, namely the LUX upper-limit ellipse in Fig.~\ref{cicj} (top-left). The boundary of the dark region follows the black line in Fig.~\ref{c1c3}, and is itself an ellipse in the $c_1^0$--$c_3^0$ plane. The correlation between $c_4^0$ and $c_6^0$ is not visible in Fig.~\ref{10cs_corr} because the {\sffamily Multinest} analysis is restricted to positive values of the $c_i^0$. It is comforting that the semi-analytic considerations at the beginning of this section and the full  {\sffamily Multinest} analysis give the same correlation pattern for the $c_i^0$'s.

\begin{figure}[t]
\begin{center}
\begin{minipage}[t]{0.49\linewidth}
\centering
\includegraphics[width=0.89\textwidth]{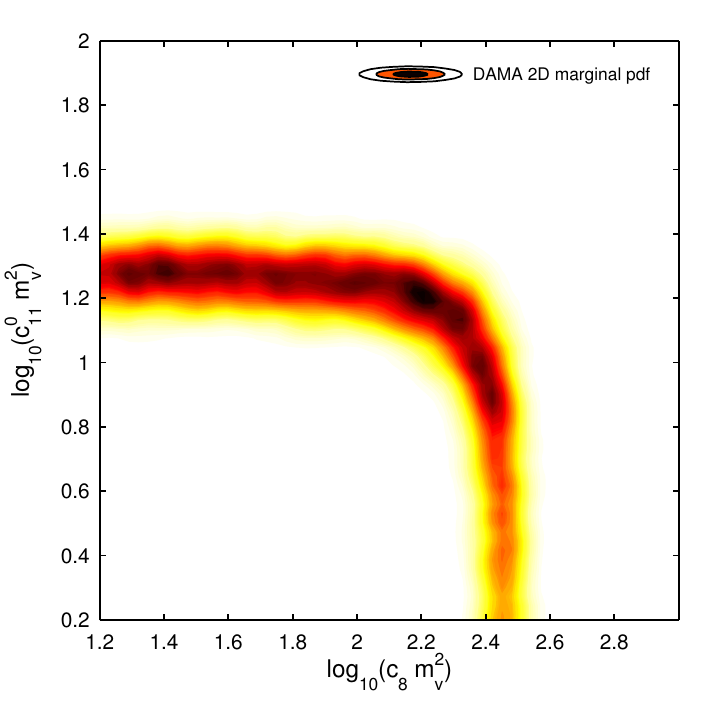}
\end{minipage}
\begin{minipage}[t]{0.49\linewidth}
\centering
\includegraphics[width=\textwidth]{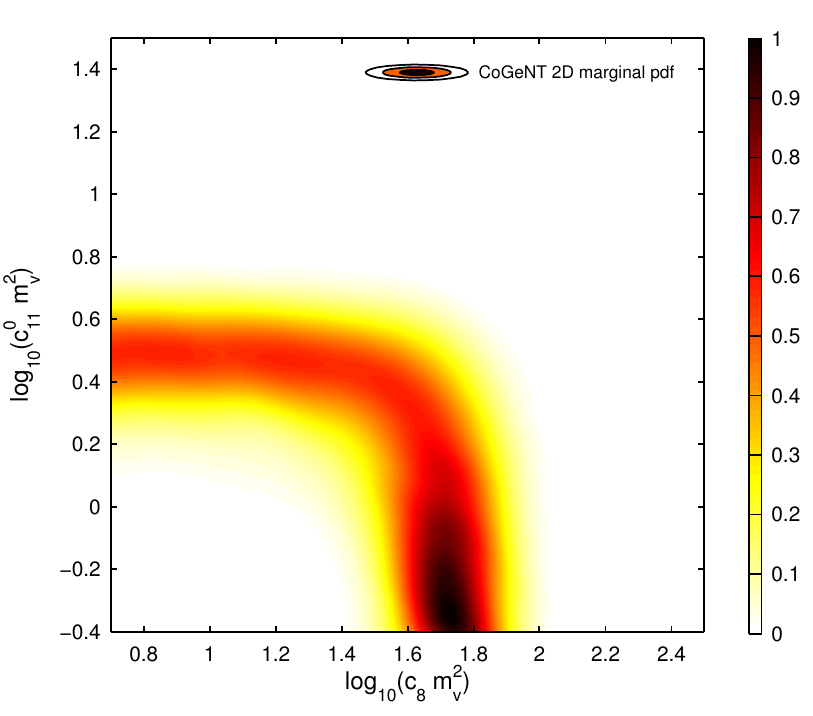}
\end{minipage}
\end{center}
\caption{2D marginal posterior PDF in the plane $c^0_{8}$ vs $c^0_{11}$ extracted from a fit of the DAMA data (left panel, scattering off Na only) and of the latest CoGeNT data (right panel) in which we have simultaneously varied $c^0_{8}$, $c^0_{11}$, $m_{\chi}$ and the nuisance parameters. Contrary to the case in which no signal is present in the data, in this case there is a clear degeneracy between the two effective couplings $c_8^0$ and $c_{11}^0$.}
\label{DamCo_corr}
\end{figure}
An interesting and important result is summarized in Fig.~\ref{10cs}, which shows the 2D marginal PDFs and the 2D profile likelihoods from our global analysis of the direct detection data in Sec.~\ref{datasets} (except DAMA and CoGeNT). In this figure we can recognize all the effects discussed in detail in the case of the LUX experiment in Fig.~\ref{c1_LUX}: the 2D marginal posterior PDFs peak at low masses because of volume effects, whereas the 2D profile likelihoods are approximately flat down to 20 GeV or so, and then start decreasing below this mass because of threshold effects. Fig.~\ref{10cs} answers the question of which is the maximum strength allowed by current direct detection data for the 10 types of interaction considered in this paper. The interactions that are currently better constrained are those described by the operators $\mathcal{O}_1$, $\mathcal{O}_3$, $\mathcal{O}_4$, $\mathcal{O}_8$ and $\mathcal{O}_{11}$. 

The results of this section show that only limits on the coupling constants $c_i^0$ derived within the profile likelihood approach are robust and physically relevant. Using present direct detection data, the Bayesian approach is unavoidably affected by volume effects generated by the marginalization process.

Our limits on $c^0_1$ and $c^0_4$ can be translated into limits on $\sigma^{\rm SI}_{N}$ and $\sigma^{\rm SD}_{N}$, respectively. For instance, for $m_{\chi}\simeq 35$~GeV, we can exclude spin-independent cross sections larger than $\sigma^{\rm SI}_{N}\simeq 1.5\times10^{-45}$~cm$^{2}$ at the 95\% CL. For $m_{\chi}\simeq 50$~GeV, we can exclude spin-dependent cross sections larger than $\sigma^{\rm SD}_{N}\simeq 2.2\times10^{-40}$~cm$^{2}$ at the 95\% CL. Our {\it global} 95\% CL limit on $\sigma^{\rm SI}_{N}$ is slightly less stringent than the one obtained by LUX, which excludes values of $\sigma^{\rm SI}_{N}$ larger than $7.6 \times 10^{-46}$~cm$^2$ at the 90\% CL, for $m_{\chi}\simeq 33$~GeV~\cite{Akerib:2013tjd}. Our limit on $\sigma^{\rm SD}_{N}$ is comparable to the one found by the XENON100 collaboration, which for $m_{\chi}\simeq 45$~GeV can exclude spin-dependent dark matter-neutron scattering cross sections larger than $3.5\times10^{-40}$~cm$^{2}$ at the 90\% CL~\cite{Aprile:2013doa}.

\section{Analyzing a signal: DAMA \& CoGeNT}
This last section is devoted to a Bayesian analysis of the DAMA and CoGeNT data. Contrary to the analysis illustrated in the previous section, we now concentrate on the interpretation of two candidate dark matter signals. We approach this problem within the theoretical and statistical frameworks introduced in Secs.~\ref{theory} and \ref{statistics}, respectively, focusing on the operators $\mathcal{O}_8$ and $\mathcal{O}_{11}$ only, and leaving the general case for future work. Here we focus on the operators $\mathcal{O}_8$ and $\mathcal{O}_{11}$, since they are among the most constrained by present direct detection data, as already mentioned above.

\label{result2}

We start with an analysis of possible degeneracies between different coupling constants. Two coupling constants are degenerate when they produce direct detection signals which cannot be experimentally disentangled. The parameters $c^0_8$ and $c^0_{11}$ are not correlated, since their correlation coefficient is zero. They are however degenerate in an analysis of the CoGeNT and DAMA results, as we will see below.

Fig.~\ref{DamCo_corr} shows the 2D marginal posterior PDFs in the $c_8^0$--$c_{11}^0$ plane extracted from an analysis of the DAMA-Na data (left panel, considering scattering off Na only, which is acceptable for $m_\chi \lesssim 20$ GeV) and CoGeNT data (right panel). In these analyses the free parameters are $m_{\chi}$,  $c_8^0$ and $c_{11}^0$, plus the relevant nuisance parameters described in Tab.~\ref{prior}. Despite the absence of a $c_8^0 c_{11}^0$ cross term in $\mu_S(m_\chi,\mathbf{c},\bfeta)$, in the case of DAMA-Na and CoGeNT we find a clear degeneracy between the parameters $c_8^0$ and $c_{11}^0$. This degeneracy has a simple origin: at small $c_{11}^0$, the DAMA-Na results can be fitted with $c_{8}^0  m_v^2 \sim 10^{2.45}$ and  $c_{11}^0  m_v^2 \lesssim 10^{0.6}$; at small  $c_{8}^0$, they can be fitted with $c_{11}^0  m_v^2 \sim 10^{1.3}$ and  $c_{8}^0  m_v^2 \lesssim 10^{1.8}$. Intermediate values of $c_8^0$ and $c_{11}^0$ can also be fitted to the DAMA-Na results, because the expected number of dark matter events is a linear combination of $(c_8^0)^2$ and $(c_{11}^0)^2$.
In the case of CoGeNT, the two limiting solutions are: (1) $c_{8}^0 m_v^2 \sim 10^{1.7}$ and $c_{11}^0 m_v^2 \lesssim 10^{0}$; (2) $c_{11}^0 m_v^2 \sim 10^{0.5}$ and $c_{8}^0 m_v^2 \lesssim10^{1.2}$. 

\begin{figure}[t]
\begin{center}
\begin{minipage}[t]{0.49\linewidth}
\centering
\includegraphics[width=\textwidth]{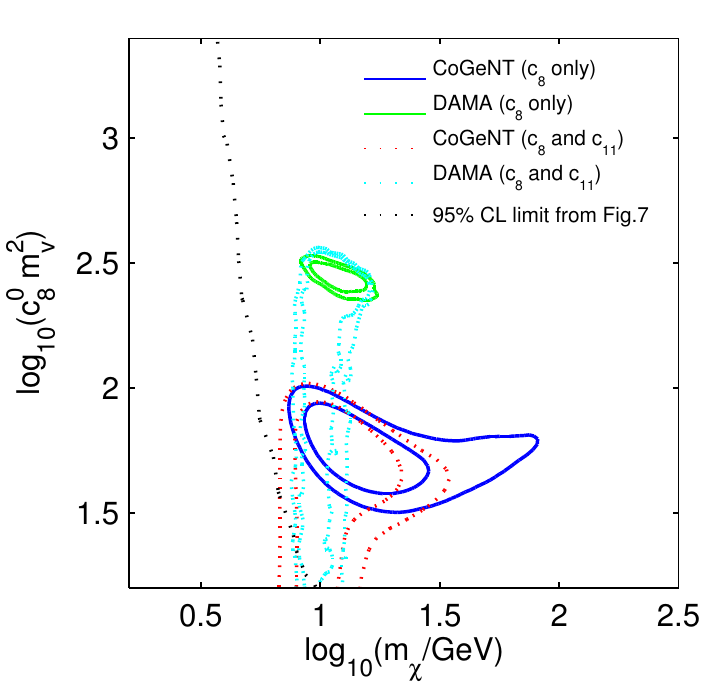}
\end{minipage}
\begin{minipage}[t]{0.49\linewidth}
\centering
\includegraphics[width=\textwidth]{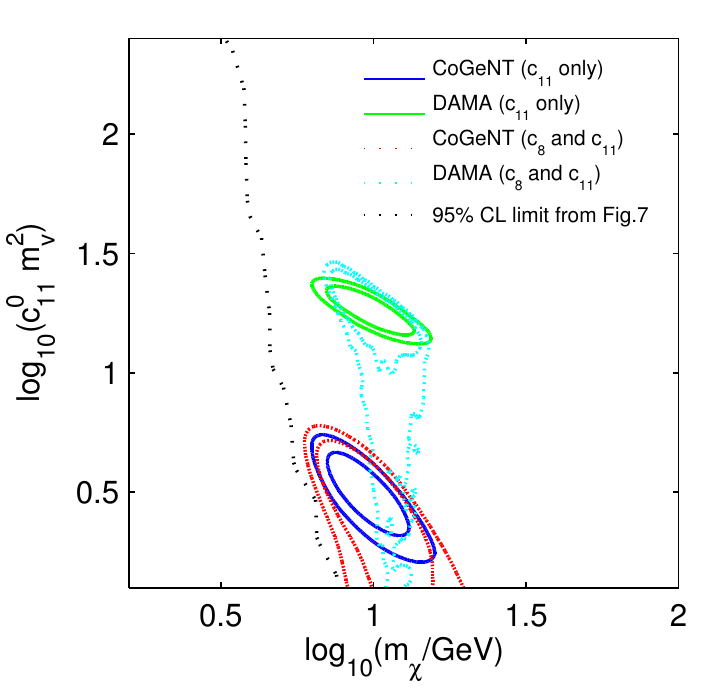}
\end{minipage}
\end{center}
\caption{68\% and 90\% credible regions in the planes $m_{\chi}$ vs $c^0_8$ (left panel) and $m_{\chi}$ vs $c^0_{11}$ (right panel) resulting from three independent analyses: (1) A fit of the DAMA data (Na only, solid green lines in the left panel)  and CoGeNT data (solid blue lines in the left panel) performed varying $m_{\chi}$ and $c^0_8$ only. (2) A fit of the DAMA data (Na only, solid green lines in the right panel)  and CoGeNT data (solid blue lines in the right panel) performed varying $m_{\chi}$ and $c^0_{11}$ only. (3) A fit of the DAMA data (Na only, dotted cyan lines)  and CoGeNT data (dotted red lines) performed varying $m_{\chi}$, $c^0_8$ and $c^0_{11}$ simultaneously (together with the nuisance parameters). When allowing both $c_8$ and $c_{11}$ to vary simultaneously, long tails appear in the marginalized posterior PDF toward small values of one of the couplings.}
\label{DamCo_fit}
\end{figure}

Fig.~\ref{DamCo_fit} shows the 2D marginal posterior PDFs in the planes $m_\chi$ vs $c_8^0$ and $m_\chi$ vs $c_{11}^0$ obtained from the same analysis of the DAMA-Na and CoGeNT data described above, i.e., varying $m_{\chi}$, $c^0_8$ and $c^0_{11}$ simultaneously  (dotted contours). The associated 68\% and 90\% CR contours are characterized by long tails extending toward the direction of zero coupling constants. These tails are related to the existence of the two limiting solutions   to the problem of fitting the data described in the previous paragraph. In addition, Fig.~\ref{DamCo_fit} also shows the 2D marginal posterior PDFs resulting from an analysis of the DAMA-Na and CoGeNT data where we have separately considered as free parameters either $c_{8}^0$ only or $c_{11}^0$ only (solid contours). The solid contours are at the end of the dotted regions.

In this example, the regions favored by DAMA-Na and CoGeNT in the plane dark matter mass vs interaction strength are well separated, discouraging therefore a global fit of the two datasets. An analysis of the remaining couplings is left for future work.

\begin{figure}[t]
\begin{center}
\begin{minipage}[t]{0.49\linewidth}
\centering
\includegraphics[width=0.9\textwidth]{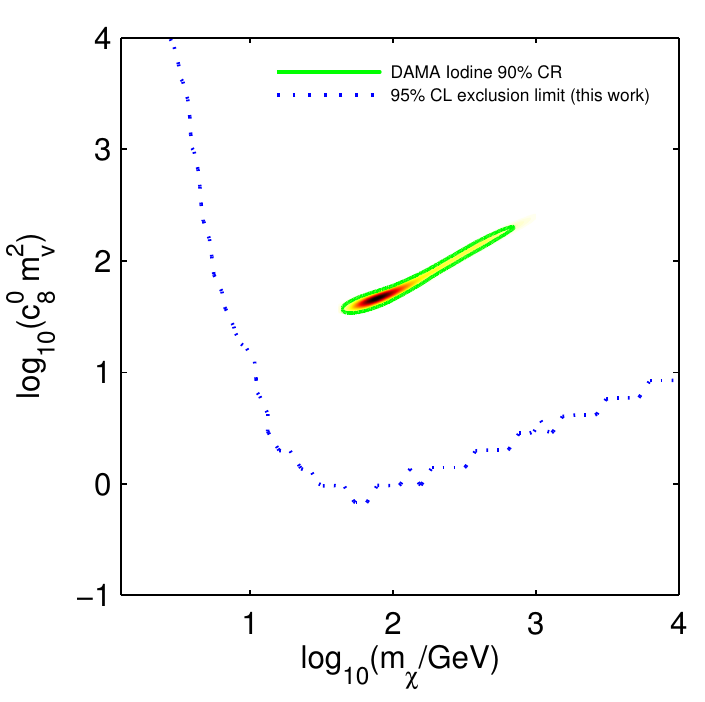}
\end{minipage}
\begin{minipage}[t]{0.49\linewidth}
\centering
\includegraphics[width=\textwidth]{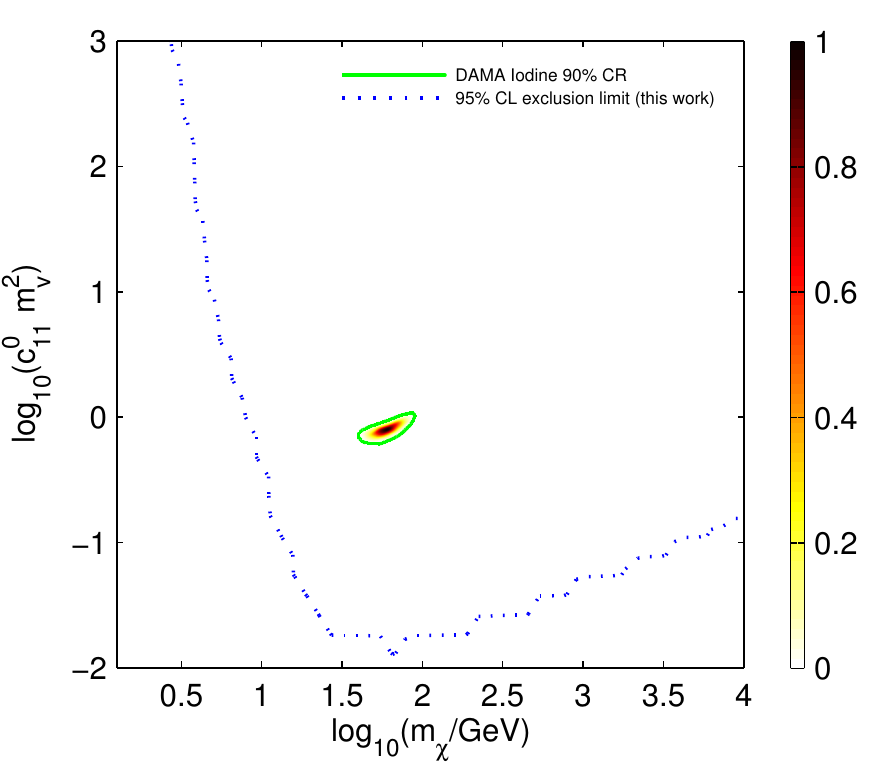}
\end{minipage}
\end{center}
\caption{Fits of the DAMA data performed varying a single coupling and $m_{\chi}$. We have assumed dark matter scattering on iodine only. We plot the resulting 2D marginal posterior PDF in the planes $m_{\chi}$ vs interaction strength for two operators studied in this paper. In both cases the interpretation of the DAMA results in terms of dark matter scattering on iodine is ruled out.}
\label{DamaI}
\end{figure}  
We conclude this section investigating whether the annual modulation signal observed by DAMA can be due to dark matter scattering on iodine. Also in this case we focus on the operators $\mathcal{O}_8$ and $\mathcal{O}_{11}$ as an illustrative example. Fig.~\ref{DamaI} shows the results of two fits performed varying either $c^0_{8}$ and $m_{\chi}$ (left panel) or $c^0_{11}$ and $m_{\chi}$ (right panel). Comparing the resulting 2D marginal posterior PDFs with our exclusion limits, we find that the DAMA annual modulation signal cannot be ascribed to dark matter interactions with iodine nuclei described by the operators $\mathcal{O}_8$ and $\mathcal{O}_{11}$.

\section{Conclusion}
\label{conclusions}
In this paper we have presented the first comprehensive analysis of the dark matter-nucleon effective interactions where the coupling constants, the dark matter mass, and additional nuisance parameters, are simultaneously considered as free parameters. To study experimental constraints on the multidimensional parameter space of coupling constants, we have implemented a Bayesian and a frequentist approach to extract credible and confidence regions from a varied sample of complementary direct detection data, including the recent LUX, CDMSlite and SuperCDMS results. We have extracted an upper bound on the 10 coupling constants characterizing the theory of heavy spin-0 and spin-1 mediators as a function of the dark matter mass, and in the limit of isospin-conserving interactions. We have calculated the posterior PDF and the profile likelihood of the model parameters, and shown credible regions and confidence levels in the planes dark matter mass vs interaction strength, marginalizing (in the Bayesian approach) and profiling out (in the frequentist approach) the uncertain or irrelevant model parameters. For the still limited experimental data currently available, the Bayesian and frequentist methods turn out to be complementary statistical indicators, in the sense that the Bayesian method is faster but subject to artificial volume effects that depend on the prior, while the frequentist method is free of volume effects but is computationally slower. 

We find that present direct detection data contain sufficient information to simultaneously constrain not only the familiar velocity-and-momentum-independent interactions (i.e., the spin-independent operator $\mathcal{O}_1=1_\chi 1_N$ and the spin-dependent operator $\mathcal{O}_4=\vec{S}_\chi \cdot \vec{S}_N$), but also the remaining velocity and momentum dependent couplings predicted by the dark matter-nucleon effective theory. The interaction most severely constrained by the current data is $\mathcal{O}_1$, followed by the interaction $\mathcal{O}_{11}$ and then $\mathcal{O}_3$, $\mathcal{O}_4$, and $\mathcal{O}_8$  (see Fig.~\ref{10cs}). Notice that the relatively strong constraints on $\mathcal{O}_{11}$ have been observed in \cite{Fan:2010gt} and indirectly in \cite{DelNobile:2013sia} (through their relativistic operator $\mathcal{O}_2$), but have not been considered in other studies \cite{Gresham:2013mua}.

In addition, we have found that strong correlations exist between $c_1^0$ and $c_3^0$ and between $c_4^0$ and $c_6^0$, associated with the interaction operators $\mathcal{O}_1 = 1_{\chi} 1_{N}$, $\mathcal{O}_3 = -i\vec{S}_N\cdot\vec{q}\times\vec{v}^{\perp}/m_N$ and $\mathcal{O}_4 = \vec{S}_{\chi}\cdot \vec{S}_{N}$, $\mathcal{O}_6 = \vec{S}_\chi\cdot\vec{q}\,\vec{S}_N\cdot\vec{q}/m_N^2$. Other correlations between the $c_i^0$'s are negligible either because there is no interference term between the corresponding operators or because they are suppressed by the smallness of the nuclear structure functions and/or momentum transfer. 

Presenting our results we have also described the difficulties found when exploring the effective-theory parameter space of large dimensionality. For instance, we have found that the marginalization process introduces important volume effects, which significantly alter the shape of the resulting 2D marginal posterior PDFs. It is therefore important, in order to assess reliable upper limits on the strength of the dark matter interactions, to calculate the profile likelihood as well, though this calculation is in general computationally very demanding and in our case it has required  about 3 million likelihood evaluations. 

In a last part of the paper we have studied two candidate dark matter signals, namely those reported by the DAMA and CoGeNT collaborations. We have approached this problem within the same theoretical and statistical frameworks used in the study of the exclusion limits. In this analysis we have considered two interaction types only for DAMA and CoGeNT, leaving the general case for future work. Analyzing a dark matter signal by varying two coupling constants and the dark matter mass, degeneracies between different couplings are apparent. They are associated with the existence of distinct solutions to the problem of fitting the data. For the two interactions types considered in this study, the regions favored by CoGeNT and DAMA in the plane dark matter mass vs interaction strength are well separated, discouraging therefore a global fit of the two datasets performed within this setup. 

In summary, we have proposed a systematic approach to the study of dark matter-nucleon effective interactions. Our approach is based on the calculation of the posterior PDF and/or the profile likelihood in the full multi-dimensional parameter space characterizing the theoretical framework. This strategy allows the extraction of global limits on the model parameters accounting for a variety of theoretical and experimental uncertainties through the introduction of nuisance parameters, which are then marginalized or profiled out during the calculation. In addition, this approach allows an interpretation of the direct detection data which is not biased by having assumed a priori the form of the dark matter-nucleon interaction. We are confident that this general and flexible approach to the analysis of dark matter direct detection data will be particularly fruitful to exploit the results of the next generation of direct detection experiments.    

\acknowledgments We  are grateful to NORDITA for support and hospitality during the workshop ``What is dark matter?'', where this work has been concluded. We also thank Eugenio Del Nobile for useful comments on a first version of this paper. R.C. acknowledges partial support from the European Union FP7 ITN INVISIBLES (Marie Curie Actions, PITN-GA-2011-289442). P.G. has been partially supported by the NSF grant PHY-1068111. 

\begin{figure}[t]
\begin{center}
\begin{minipage}[t]{0.49\linewidth}
\centering
\includegraphics[width=\textwidth]{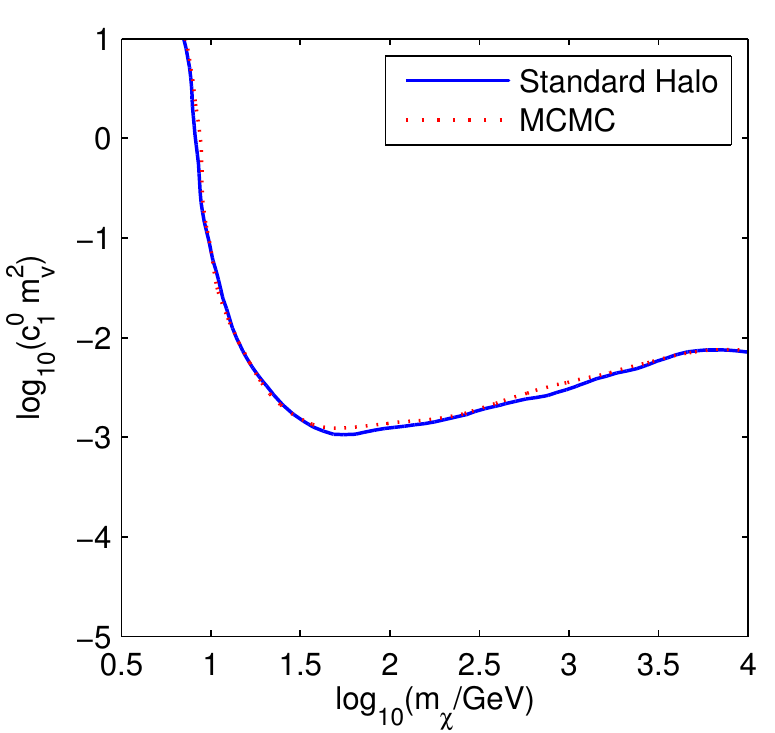}
\end{minipage}
\begin{minipage}[t]{0.49\linewidth}
\centering
\includegraphics[width=\textwidth]{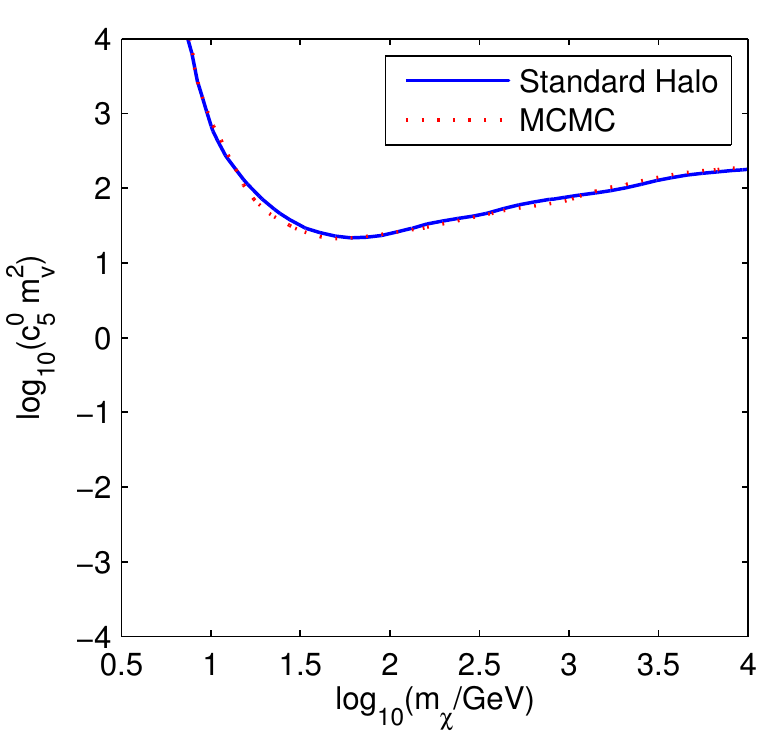}
\end{minipage}
\end{center}
\caption{99\% credible regions in the planes $m_{\chi}$ vs $c^0_1$ (left panel) and $m_{\chi}$ vs $c^0_{5}$ (right panel) resulting from an analysis of the XENON100 data assuming the galactic model of Ref.~\cite{Bozorgnia:2013pua}, instead of the standard dark matter halo (in this analysis, the coupling constants not shown in the figures are set to zero, and the PDFs are marginalized over the 8 astrophysical model parameters). Changing the astrophysical assumptions modifies the 2D marginal posterior PDFs only moderately.}
\label{astro}
\end{figure}  

\appendix
\section{Changing the astrophysical assumptions}
\label{astromcmc}
In this paper we have assumed the standard dark matter halo to extract limits on the couplings $c_i^0$ from present direct detection data. We check here with two examples to which extent our conclusions would have been affected by having assumed a different astrophysical configuration. In this appendix, we assume the galactic model studied in depth in Refs.~\cite{Catena:2009mf,Catena:2011kv, Bozorgnia:2013pua}. It features 8 parameters describing a galactic bulge, a stellar disk and a spherical dark matter halo. This model~\cite{Bozorgnia:2013pua} has been generalized to include an anisotropic velocity distribution for the Milky Way dark matter particles.

In this appendix we calculate the 2D marginal posterior PDFs in the planes $m_\chi$ vs  $c_1^0$ and $m_\chi$ vs  $c_5^0$ associated with the XENON100 data assuming the more complex astrophysical configuration in~\cite{Bozorgnia:2013pua} and marginalizing over its 8 astrophysical parameters. For these parameters we assume the prior PDFs shown in Fig.~4 of Ref.~\cite{Bozorgnia:2013pua}. Having computed the 2D marginal posterior PDFs, we then compare the corresponding 99\% CR contours with those obtained assuming a standard dark matter halo. The results of this analysis are shown in Fig.~\ref{astro}. The left panel refers to the operator $\mathcal{O}_1$, as an example of velocity/momentum independent operator, whereas the right panel refers to the operator $\mathcal{O}_5$, which is instead a velocity/momentum dependent operator. As one can see from this figure, changing the astrophysical assumptions modifies the 2D marginal posterior PDFs only moderately. Our interpretation of present direct detection data is more sensitive to the assumptions made regarding the underlying dark matter-nucleon interaction. Notice however that the mean galactic model found in Ref.~\cite{Bozorgnia:2013pua} and the standard dark matter halo have very similar local dark matter densities. A more drastic modification of the local density would have induced more significant changes in the 2D marginal posterior PDFs reported in Fig.~\ref{astro}.

\section{Dark matter response functions}
\label{dmrfun}
 In the following we list the dark matter response functions used in the calculations presented in this paper. These have been obtained from the ones derived in Ref.~\cite{Anand:2013yka} setting to zero the couplings $c^{\tau}_{12},\dots,c^{\tau}_{15}$, with $\tau=0,1$:

\allowdisplaybreaks
\begin{eqnarray}
\label{eq:Rfunctions}
 R_{M}^{\tau \tau^\prime}({v}^{\perp 2}_{\chi T}, {{q}^{2} \over m_N^2}) &=& c_1^\tau c_1^{\tau^\prime } + {j_\chi (j_\chi+1) \over 3} \left[ {{q}^{2} \over m_N^2} {v}^{\perp 2}_{\chi T} c_5^\tau c_5^{\tau^\prime }+{v}^{\perp 2}_{\chi T} c_8^\tau c_8^{\tau^\prime }
+ {{q}^{2} \over m_N^2} c_{11}^\tau c_{11}^{\tau^\prime } \right] \nonumber \\
 R_{\Phi^{\prime \prime}}^{\tau \tau^\prime}({v}^{\perp 2}_{\chi T}, {{q}^{2} \over m_N^2}) &=& {{q}^{2} \over 4 m_N^2} c_3^\tau c_3^{\tau^\prime } 
 \nonumber \\
 R_{\Phi^{\prime \prime} M}^{\tau \tau^\prime}({v}^{\perp 2}_{\chi T}, {{q}^{2} \over m_N^2}) &=&  c_3^\tau c_1^{\tau^\prime } 
 \nonumber \\
  R_{\tilde{\Phi}^\prime}^{\tau \tau^\prime}({v}^{\perp 2}_{\chi T}, {{q}^{2} \over m_N^2}) &=& 0
  \nonumber \\
   R_{\Sigma^{\prime \prime}}^{\tau \tau^\prime}({v}^{\perp 2}_{\chi T}, {{q}^{2} \over m_N^2})  &=&{{q}^{2} \over 4 m_N^2} c_{10}^\tau  c_{10}^{\tau^\prime } +
  {j_\chi (j_\chi+1) \over 12} \left[ c_4^\tau c_4^{\tau^\prime} + 
 {{q}^{2} \over m_N^2} ( c_4^\tau c_6^{\tau^\prime }+c_6^\tau c_4^{\tau^\prime })+
 {{q}^{4} \over m_N^4} c_{6}^\tau c_{6}^{\tau^\prime } \right] \nonumber \\
    R_{\Sigma^\prime}^{\tau \tau^\prime}({v}^{\perp 2}_{\chi T}, {{q}^{2} \over m_N^2})  &=&{1 \over 8} \left[ {{q}^{2} \over  m_N^2}  {v}^{\perp 2}_{\chi T} c_{3}^\tau  c_{3}^{\tau^\prime } + {v}^{\perp 2}_{\chi T}  c_{7}^\tau  c_{7}^{\tau^\prime }  \right]
       + {j_\chi (j_\chi+1) \over 12} \left[ c_4^\tau c_4^{\tau^\prime} + 
        {{q}^{2} \over m_N^2} c_9^\tau c_9^{\tau^\prime }
        \right] \nonumber \\
     R_{\Delta}^{\tau \tau^\prime}({v}^{\perp 2}_{\chi T}, {{q}^{2} \over m_N^2})&=&  {j_\chi (j_\chi+1) \over 3} \left[ {{q}^{\,2} \over m_N^2} c_{5}^\tau c_{5}^{\tau^\prime }+ c_{8}^\tau c_{8}^{\tau^\prime } \right] \nonumber \\
 R_{\Delta \Sigma^\prime}^{\tau \tau^\prime}({v}^{\perp 2}_{\chi T}, {{q}^{2} \over m_N^2})&=& {j_\chi (j_\chi+1) \over 3} \left[c_{5}^\tau c_{4}^{\tau^\prime }-c_8^\tau c_9^{\tau^\prime} \right].
\end{eqnarray}

\providecommand{\newblock}{}


\begin{thebibliography}{10}
\expandafter\ifx\csname url\endcsname\relax
  \def\url#1{{\tt #1}}\fi
\expandafter\ifx\csname urlprefix\endcsname\relax\def\urlprefix{URL }\fi
\providecommand{\eprint}[2][]{\url{#2}}

\bibitem{Ade:2013zuv}
Ade P {\em et~al.\/} (Planck Collaboration) 2013  (\textit{Preprint}
  \eprint{1303.5076})

\bibitem{Zwicky}
Zwicky F 1933 {\em Phys. Acta\/} {\bf 6} 110--127

\bibitem{Kolb:1990vq}
Kolb E~W and Turner M~S 1990 {\em Front.Phys.\/} {\bf 69} 1--547

\bibitem{Jungman:1995df}
Jungman G, Kamionkowski M and Griest K 1996 {\em Phys.Rept.\/} {\bf 267}
  195--373 (\textit{Preprint} \eprint{hep-ph/9506380})

\bibitem{Bertone:2004pz}
Bertone G, Hooper D and Silk J 2005 {\em Phys.Rept.\/} {\bf 405} 279--390
  (\textit{Preprint} \eprint{hep-ph/0404175})

\bibitem{Kuhlen:2012ft}
Kuhlen M, Vogelsberger M and Angulo R 2012 {\em Phys.Dark Univ.\/} {\bf 1}
  50--93 (\textit{Preprint} \eprint{1209.5745})

\bibitem{Strigari:2013iaa}
Strigari L~E 2013 {\em Phys.Rept.\/} {\bf 531} 1--88 (\textit{Preprint}
  \eprint{1211.7090})

\bibitem{Goodman:1984dc}
Goodman M~W and Witten E 1985 {\em Phys.Rev.\/} {\bf D31} 3059

\bibitem{Baudis:2012ig}
Baudis L 2012 {\em Phys.Dark Univ.\/} {\bf 1} 94--108 (\textit{Preprint}
  \eprint{1211.7222})

\bibitem{Cerdeno:2010jj}
Cerdeno D~G and Green A~M 2010  (\textit{Preprint} \eprint{1002.1912})

\bibitem{DelNobile:2013gba}
Del~Nobile E, Gelmini G~B, Gondolo P and Huh J~H 2013  (\textit{Preprint}
  \eprint{1311.4247})

\bibitem{Frandsen:2013cna}
Frandsen M~T, Kahlhoefer F, McCabe C, Sarkar S and Schmidt-Hoberg K 2013 {\em
  JCAP\/} {\bf 1307} 023 (\textit{Preprint} \eprint{1304.6066})

\bibitem{Schwetz:2011xm}
Schwetz T and Zupan J 2011 {\em JCAP\/} {\bf 1108} 008 (\textit{Preprint}
  \eprint{1106.6241})

\bibitem{Farina:2011pw}
Farina M, Pappadopulo D, Strumia A and Volansky T 2011 {\em JCAP\/} {\bf 1111}
  010 (\textit{Preprint} \eprint{1107.0715})

\bibitem{Kopp:2009qt}
Kopp J, Schwetz T and Zupan J 2010 {\em JCAP\/} {\bf 1002} 014
  (\textit{Preprint} \eprint{0912.4264})

\bibitem{Savage:2010tg}
Savage C, Gelmini G, Gondolo P and Freese K 2011 {\em Phys.Rev.\/} {\bf D83}
  055002 (\textit{Preprint} \eprint{1006.0972})

\bibitem{Baltz:2004aw}
Baltz E~A and Gondolo P 2004 {\em JHEP\/} {\bf 0410} 052 (\textit{Preprint}
  \eprint{hep-ph/0407039})

\bibitem{Ellis:2005mb}
Ellis J~R, Olive K~A, Santoso Y and Spanos V~C 2005 {\em Phys.Rev.\/} {\bf D71}
  095007 (\textit{Preprint} \eprint{hep-ph/0502001})

\bibitem{Catena:2013pka}
Catena R and Covi L 2013  (\textit{Preprint} \eprint{1310.4776})

\bibitem{Chang:2009yt}
Chang S, Pierce A and Weiner N 2010 {\em JCAP\/} {\bf 1001} 006
  (\textit{Preprint} \eprint{0908.3192})

\bibitem{Feng:2011vu}
Feng J~L, Kumar J, Marfatia D and Sanford D 2011 {\em Phys.Lett.\/} {\bf B703}
  124--127 (\textit{Preprint} \eprint{1102.4331})

\bibitem{Fornengo:2011sz}
Fornengo N, Panci P and Regis M 2011 {\em Phys.Rev.\/} {\bf D84} 115002
  (\textit{Preprint} \eprint{1108.4661})

\bibitem{DelNobile:2014eta}
Del~Nobile E, Gelmini G~B, Gondolo P and Huh J~H 2014  (\textit{Preprint}
  \eprint{1401.4508})

\bibitem{Gresham:2013mua}
Gresham M~I and Zurek K~M 2014 {\em Phys.Rev.\/} {\bf D89} 016017
  (\textit{Preprint} \eprint{1311.2082})

\bibitem{Liang:2013dsa}
Liang Z~L and Wu Y~L 2014 {\em Phys.Rev.\/} {\bf D89} 013010 (\textit{Preprint}
  \eprint{1308.5897})

\bibitem{deSimone:2014pda}
De~Simone A, Giudice G~F and Strumia A 2014  (\textit{Preprint}
  \eprint{1402.6287})

\bibitem{Peter:2013aha}
Peter A~H~G, Gluscevic V, Green A~M, Kavanagh B~J and Lee S~K 2013
  (\textit{Preprint} \eprint{1310.7039})

\bibitem{DelNobile:2013cva}
Del~Nobile E, Gelmini G, Gondolo P and Huh J~H 2013 {\em JCAP\/} {\bf 1310} 048
  (\textit{Preprint} \eprint{1306.5273})
  
 \bibitem{Cherry:2014wia}
 Cherry~J~F, Frandsen~M~T and Shoemaker~I~M  (\textit{Preprint}  
  \eprint{1405.1420})

\bibitem{Fitzpatrick:2012ib}
Fitzpatrick A~L, Haxton W, Katz E, Lubbers N and Xu Y 2012  (\textit{Preprint}
  \eprint{1211.2818})

\bibitem{Fitzpatrick:2012ix}
Fitzpatrick A~L, Haxton W, Katz E, Lubbers N and Xu Y 2013 {\em JCAP\/} {\bf
  1302} 004 (\textit{Preprint} \eprint{1203.3542})

\bibitem{Anand:2013yka}
Anand N, Fitzpatrick A~L and Haxton W 2013  (\textit{Preprint}
  \eprint{1308.6288})

\bibitem{Fan:2010gt}
Fan J, Reece M and Wang L~T 2010 {\em JCAP\/} {\bf 1011} 042 (\textit{Preprint}
  \eprint{1008.1591})

\bibitem{DelNobile:2013sia}
Cirelli M, Del~Nobile E and Panci P 2013 {\em JCAP\/} {\bf 1310} 019
  (\textit{Preprint} \eprint{1307.5955})

\bibitem{Panci:2014gga}
 Panci~P {\em Adv.~High Energy Phys.}  {\bf 2014} (2014) 681312
  (\textit{Preprint} \eprint{1402.1507})

\bibitem{Gresham:2014vja}
Gresham M~I and Zurek K~M 2014  (\textit{Preprint} \eprint{1401.3739})

\bibitem{Buckley:2013jwa}
Buckley M~R 2013 {\em Phys.Rev.\/} {\bf D88} 055028 (\textit{Preprint}
  \eprint{1308.4146})

\bibitem{Cirigliano:2013zta}
Cirigliano~V, Graesser~M~L, Ovanesyan~G and Shoemaker~I~M (\textit{Preprint}
  \eprint{1311.5886})

\bibitem{Bozorgnia:2013pua}
Bozorgnia N, Catena R and Schwetz T 2013 {\em JCAP\/} {\bf 1312} 050
  (\textit{Preprint} \eprint{1310.0468})

\bibitem{Trotta:2006ew}
Trotta R, de~Austri R~R and Roszkowski L 2007 {\em New Astron. Rev.\/} {\bf 51}
  316--320 (\textit{Preprint} \eprint{astro-ph/0609126})

\bibitem{Akrami:2010dn}
Akrami Y, Savage C, Scott P, Conrad J and Edsjo J 2011 {\em JCAP\/} {\bf 1104}
  012 (\textit{Preprint} \eprint{1011.4318})

\bibitem{Akrami:2010cz}
Akrami Y, Savage C, Scott P, Conrad J and Edsjo J 2011 {\em JCAP\/} {\bf 1107}
  002 (\textit{Preprint} \eprint{1011.4297})

\bibitem{Pato:2010zk}
Pato M, Baudis L, Bertone G, Ruiz~de Austri R, Strigari L~E {\em et~al.\/} 2011
  {\em Phys.Rev.\/} {\bf D83} 083505 (\textit{Preprint} \eprint{1012.3458})

\bibitem{Arina:2011si}
Arina C, Hamann J and Wong Y~Y 2011 {\em JCAP\/} {\bf 1109} 022
  (\textit{Preprint} \eprint{1105.5121})

\bibitem{Arina:2011zh}
Arina C, Hamann J, Trotta R and Wong Y~Y 2012 {\em JCAP\/} {\bf 1203} 008
  (\textit{Preprint} \eprint{1111.3238})

\bibitem{Peter:2011eu}
Peter A~H 2011 {\em Phys.Rev.\/} {\bf D83} 125029 (\textit{Preprint}
  \eprint{1103.5145})

\bibitem{Arina:2012dr}
Arina C 2012 {\em Phys.Rev.\/} {\bf D86} 123527 (\textit{Preprint}
  \eprint{1210.4011})

\bibitem{Strege:2012kv}
Strege C, Trotta R, Bertone G, Peter A~H and Scott P 2012 {\em Phys.Rev.\/}
  {\bf D86} 023507 (\textit{Preprint} \eprint{1201.3631})

\bibitem{Arina:2013jma}
Arina C 2013  (\textit{Preprint} \eprint{1310.5718})

\bibitem{Cerdeno:2013gqa}
Cerde–o D, Cuesta C, Fornasa M, Garc'a E, Ginestra C {\em et~al.\/} 2013 {\em
  JCAP\/} {\bf 1307} 028 (\textit{Preprint} \eprint{1304.1758})

\bibitem{Cerdeno:2014uga}
Cerdeno D, Cuesta C, Fornasa M, Garcia E, Ginestra C {\em et~al.\/} 2014
  (\textit{Preprint} \eprint{1403.3539})
  
\bibitem{Cowan}
Cowan G 1998 {\em Oxford Universitiy Press} 1--197  

\bibitem{2012arXiv1209.3339F}
{Freese} K, {Lisanti} M and {Savage} C 2012 {\em ArXiv e-prints\/}
  (\textit{Preprint} \eprint{1209.3339})

\bibitem{2011JHEP...06..042F}
{Feroz} F, {Cranmer} K, {Hobson} M, {Ruiz de Austri} R and {Trotta} R 2011 {\em
  Journal of High Energy Physics\/} {\bf 6} 42 (\textit{Preprint}
  \eprint{1101.3296})

\bibitem{Behnke:2012ys}
Behnke E {\em et~al.\/} (COUPP Collaboration) 2012 {\em Phys.Rev.\/} {\bf D86}
  052001 (\textit{Preprint} \eprint{1204.3094})

\bibitem{Archambault:2012pm}
Archambault S {\em et~al.\/} (PICASSO Collaboration) 2012 {\em Phys.Lett.\/}
  {\bf B711} 153--161 (\textit{Preprint} \eprint{1202.1240})

\bibitem{Felizardo:2011uw}
Felizardo M, Girard T, Morlat T, Fernandes A, Ramos A {\em et~al.\/} 2012 {\em
  Phys.Rev.Lett.\/} {\bf 108} 201302 (\textit{Preprint} \eprint{1106.3014})

\bibitem{Aalseth:2014jpa}
Aalseth C, Barbeau P, Colaresi J, Leon J~D, Fast J {\em et~al.\/} 2014
  (\textit{Preprint} \eprint{1401.6234})

\bibitem{Feroz:2008xx}
Feroz F, Hobson M and Bridges M 2009 {\em Mon.Not.Roy.Astron.Soc.\/} {\bf 398}
  1601--1614 (\textit{Preprint} \eprint{0809.3437})

\bibitem{Feroz:2007kg}
Feroz F and Hobson M 2008 {\em Mon.Not.Roy.Astron.Soc.\/} {\bf 384} 449
  (\textit{Preprint} \eprint{0704.3704})

\bibitem{Feroz:2013hea}
Feroz F, Hobson M, Cameron E and Pettitt A 2013  (\textit{Preprint}
  \eprint{1306.2144})

\bibitem{Lewis:2002ah}
Lewis A and Bridle S 2002 {\em Phys. Rev.\/} {\bf D66} 103511
  (\textit{Preprint} \eprint{astro-ph/0205436})

\bibitem{Austri:2006pe}
de~Austri R~R, Trotta R and Roszkowski L 2006 {\em JHEP\/} {\bf 0605} 002
  (\textit{Preprint} \eprint{hep-ph/0602028})

\bibitem{Ahmed:2009zw}
Ahmed Z {\em et~al.\/} (CDMS-II Collaboration) 2010 {\em Science\/} {\bf 327}
  1619--1621 (\textit{Preprint} \eprint{0912.3592})

\bibitem{Ahmed:2010wy}
Ahmed Z {\em et~al.\/} (CDMS-II Collaboration) 2011 {\em Phys.Rev.Lett.\/} {\bf
  106} 131302 (\textit{Preprint} \eprint{1011.2482})

\bibitem{Agnese:2014aze}
Agnese R {\em et~al.\/} (SuperCDMS Collaboration) 2014  (\textit{Preprint}
  \eprint{1402.7137})

\bibitem{Agnese:2013jaa}
Agnese R {\em et~al.\/} (SuperCDMSSoudan Collaboration) 2014 {\em
  Phys.Rev.Lett.\/} {\bf 112} 041302 (\textit{Preprint} \eprint{1309.3259})

\bibitem{Aprile:2012nq}
Aprile E {\em et~al.\/} (XENON100 Collaboration) 2012 {\em Phys.Rev.Lett.\/}
  {\bf 109} 181301 (\textit{Preprint} \eprint{1207.5988})

\bibitem{Aprile:2011hi}
Aprile E {\em et~al.\/} (XENON100 Collaboration) 2011 {\em Phys.Rev.Lett.\/}
  {\bf 107} 131302 (\textit{Preprint} \eprint{1104.2549})

\bibitem{Aprile:2011hx}
Aprile E {\em et~al.\/} (XENON100 Collaboration) 2011 {\em Phys.Rev.\/} {\bf
  D84} 052003 (\textit{Preprint} \eprint{1103.0303})

\bibitem{Angle:2011th}
Angle J {\em et~al.\/} (XENON10 Collaboration) 2011 {\em Phys.Rev.Lett.\/} {\bf
  107} 051301 (\textit{Preprint} \eprint{1104.3088})

\bibitem{Lewin:1995rx}
Lewin J and Smith P 1996 {\em Astropart.Phys.\/} {\bf 6} 87--112

\bibitem{Akerib:2013tjd}
Akerib D {\em et~al.\/} (LUX Collaboration) 2013  (\textit{Preprint}
  \eprint{1310.8214})

\bibitem{Bernabei:2010mq}
Bernabei R {\em et~al.\/} (DAMA Collaboration, LIBRA Collaboration) 2010 {\em
  Eur.Phys.J.\/} {\bf C67} 39--49 (\textit{Preprint} \eprint{1002.1028})

\bibitem{Aalseth:2014eft}
Aalseth C {\em et~al.\/} (CoGeNT Collaboration) 2014  (\textit{Preprint}
  \eprint{1401.3295})

\bibitem{Davis:2014bla} 
Davis~J~H, McCabe~C and Boehm~C (\textit{Preprint}
  \eprint{1405.0495})

\bibitem{Aprile:2013doa}
Aprile E {\em et~al.\/} (XENON100 Collaboration) 2013 {\em Phys.Rev.Lett.\/} 
 {\bf 111} 021301 (\textit{Preprint} \eprint{1301.6620})

\bibitem{Catena:2009mf}
Catena R and Ullio P 2010 {\em JCAP\/} {\bf 1008} 004 (\textit{Preprint}
  \eprint{0907.0018})

\bibitem{Catena:2011kv}
Catena R and Ullio P 2012 {\em JCAP\/} {\bf 1205} 005 (\textit{Preprint}
  \eprint{1111.3556})

\end{thebibliography}
\end{document}